\documentclass[10pt,a4paper]{article}
\usepackage{jheppub}
\usepackage[dvipsnames]{xcolor}
\usepackage[T1]{fontenc}
\usepackage{graphicx}
\usepackage{amsmath}
\usepackage{xspace}
\usepackage[numbers,sort&compress]{natbib}
\usepackage{subfig}
\usepackage[shortlabels]{enumitem}

\usepackage{pstricks}
\usepackage{graphicx}
\usepackage{xspace}
\usepackage{hyperref}
\usepackage{lipsum}
\usepackage{multirow}

\newcommand{\pT}{\ensuremath{p_{\rm{T}}}\xspace}
\newcommand{\pTH}{\ensuremath{p_{\rm{T},H}}\xspace}
\newcommand{\pTj}{\ensuremath{p_{\rm{T},j}}\xspace}

\newcommand{\mHj}{\ensuremath{m_{Hj}}\xspace}
\newcommand{\mjj}{\ensuremath{m_{j_1j_2}}\xspace}

\newcommand{\pphj}{\ensuremath{pp \rightarrow H+j}\xspace}
\newcommand{\pphjj}{\ensuremath{pp \rightarrow H+jj}\xspace}

\newcommand{\hj}{\ensuremath{\rm{H}+}jet\xspace}
\newcommand{\hjj}{\ensuremath{\rm{H}+2\,}jets\xspace}

\newcommand\GeV{\ensuremath{\mathrm{GeV}}\xspace}
\newcommand\TeV{\ensuremath{\mathrm{TeV}}\xspace}
\newcommand\ord{\ensuremath{O}\xspace}

\newcommand{\pTHhigh}{\ensuremath{p_{\rm{T},H}> 300\,\mathrm{GeV}}\xspace}

\newcommand{\NLO}{\ensuremath{\rm{NLO}}\xspace}
\newcommand{\FTapprox}{\ensuremath{\rm{FT}_{\rm approx}}\xspace}
\newcommand{\HTL}{\ensuremath{\rm{HTL}}\xspace}

\newcommand{\nnlojet}{{\rmfamily\scshape NNLOJET}\xspace}
\newcommand{\secdec}{{\rmfamily\scshape SecDec-3}\xspace}
\newcommand{\olotter}{{\rmfamily\scshape OpenLoops2.2}\xspace}
\newcommand{\openloops}{{\rmfamily\scshape OpenLoops2}\xspace}
\newcommand{\otter}{{\rmfamily\slshape Otter}\xspace}

\def\d{{\rm d}}

\usepackage[left,mathlines]{lineno} 
\usepackage{soul} 

\def\bnote{\begin{notes}}
\def\enote{\end{notes}}
\usepackage{tcolorbox}
\newcounter{notescounter}
\newenvironment{notes}{\stepcounter{notescounter}
\begin{center} 
\begin{minipage}[t]{\textwidth} 
\begin{tcolorbox}[colback=Apricot,colframe=Apricot]
[\thenotescounter]\;}{ 
\end{tcolorbox} 
\end{minipage}  
\end{center}}

\title{Top-quark mass effects in $H$+jet and $H$+2~jets production}

\author[1,2,3]{X. Chen,}
\author[4]{A. Huss,}
\author[5]{S. P. Jones,}
\author[1,2,3]{M. Kerner,}
\author[3]{J.-N. Lang,}
\author[6]{J. M. Lindert,}
\author[3,7]{H. Zhang}

\affiliation[1]{\footnotesize Institute for Theoretical Physics, Karlsruhe Institute of Technology, 76131 Karlsruhe, Germany \normalsize}
\affiliation[2]{\footnotesize Institute for Astroparticle Physics, Karlsruhe Institute of Technology, 76344 Eggenstein-Leopoldshafen, Germany\normalsize}
\affiliation[3]{\footnotesize Physik-Institut, Universit\"at Z\"urich, CH-8057 Z\"urich, Switzerland \normalsize}
\affiliation[4]{\footnotesize Theoretical Physics Department, CERN, 1211 Geneva 23, Switzerland \normalsize}
\affiliation[5]{\footnotesize Institute for Particle Physics Phenomenology, Durham University, Durham, DH1 3LE, UK \normalsize}
\affiliation[6]{\footnotesize Department of Physics and Astronomy, University of Sussex, Brighton BN1 9QH, UK \normalsize}
\affiliation[7]{\footnotesize Institut f\"{u}r Theoretische Teilchenphysik, Karlsruhe Institute of Technology, 76128 Karlsruhe, Germany\normalsize}

\abstract{
    We present calculations of Higgs boson production via gluon-gluon fusion in
    association with one or two additional jets at next-to-leading order in QCD.
    The calculation of $H$+jet is exact in the treatment of the top-quark mass, 
    whereas for the $H$+2~jets
    calculation the two-loop virtual amplitudes are approximated via a
    reweighting with leading-order mass effects, while keeping all top-quark mass effects
    in the real radiation contributions. For $H$+jet production, this study extends a 
    previous calculation, revealing an error in the previous results.
    For total and differential cross sections, we present new results and
    compare the QCD corrections with the infinite top-mass limit, for which we
    find a strikingly good agreement if all amplitudes are rescaled by the leading-order mass dependence.
}

\begin{document}
\preprint{
\begin{flushright}
KA-TP-21-2021 \\ 
TTP21-035\\
CERN-TH-2021-153\\
P3H-21-070\\
ZU-TH 48/21\\
IPPP/21/39
\end{flushright}
}

\maketitle
\flushbottom

\section{Introduction}

The current and upcoming runs of the Large Hadron Collider (LHC) are stress-testing the Standard Model (SM) of particle physics at an unprecedented level. In this respect one of the main objectives of Run~3 
and the high-luminosity phase of the LHC (HL-LHC)  will be a further detailed investigation of the Higgs sector. The abundant future data samples will allow the range of Higgs analyses to be extended to multi-dimensional measurements and high-energy tails of kinematic distributions. A key observable in this regard is the transverse momentum distribution of the Higgs boson, $\pTH$, which serves as a unique probe of physics Beyond the Standard Model (BSM)~\cite{Arnesen:2008fb,Banfi:2013yoa,Azatov:2013xha,Grojean:2013nya,Harlander:2013oja,Dawson:2014ora,Langenegger:2015lra,Grazzini:2016paz,Deutschmann:2017qum,Battaglia:2021nys}. Given the expected experimental data sets, this distribution will be measured both inclusively and in association with jets up to several hundreds of \GeV in the Higgs transverse momentum~\cite{ATLAS:2018ibe,CMS:2018qgz}. Already now experimental measurements by ATLAS and CMS yield sensitivity up to few hundred \GeV~\cite{ATLAS:2021nsx,Sirunyan:2020hwz}.

Both in the inclusive case and for the production in association with jets, the dominant Higgs production mode in the SM originates via a top-quark loop in gluon-gluon fusion, however, at large transverse momentum eventually also vector-boson fusion (VBF), Higgsstrahlung (VH) and top-pair associated Higgs production contribute significantly~\cite{Becker:2020rjp}. For the production in association with jets,
 additional constraints on jet invariant masses and/or rapidities allow the relative fraction of the VBF
 events to be enhanced~\cite{Campanario:2018ppz,Buckley:2021gfw}. Precise VBF measurements will allow us to constrain on the one hand the electroweak (EW) couplings of the Higgs, and on the other hand when restricting to large Higgs transverse momentum they will allow for complementary constraints on models of new physics~\cite{Artoisenet:2013puc,Maltoni:2013sma,Greljo:2015sla,Degrande:2016dqg,Greljo:2017spw,Araz:2020zyh}.     
In this regard, one of the dominant uncertainties in VBF measurements originates from the background modelling of the gluon-induced Higgs production mode. 

The loop-induced nature of the gluon-gluon fusion Higgs production process makes higher-order corrections notoriously difficult to calculate. In QCD
fixed-order perturbation theory including mass effects, inclusive Higgs
production at next-to-next-to-leading order (NNLO) was calculated only very
recently~\cite{Czakon:2021yub}, Higgs plus jet production is known at
next-to-leading order
(NLO)~\cite{Jones:2018hbb,Lindert:2018iug,Neumann:2018bsx,Kerner:2019qgb}, while
Higgs plus dijet production (and beyond) is only known at leading order
(LO)~\cite{DelDuca:2001eu,DelDuca:2001fn,Neumann:2016dny,Andersen:2018kjg,Budge:2020oyl}.
In the NLO computations of Higgs plus jet production the crucial two-loop virtual contributions have been obtained numerically~\cite{Borowka:2015mxa,Borowka:2017idc} in Refs.~\cite{Jones:2018hbb,Kerner:2019qgb} respectively via a suitable high-energy expansion~\cite{Kudashkin:2017skd} in Refs.~\cite{Lindert:2018iug,Neumann:2018bsx}.

Formally, below the top-quark threshold, higher precision can be achieved via the heavy top loop (HTL)
approximation,
effectively integrating out the top-quark loop~\cite{Wilczek:1977zn}. In
the HTL approximation, inclusive Higgs production is known at N3LO~\cite{Anastasiou:2015vya,Dulat:2017prg,Cieri:2018oms,Mistlberger:2018etf,Chen:2021isd}, Higgs plus jet production at NNLO~\cite{Boughezal:2013uia,Chen:2014gva,Boughezal:2015dra,Boughezal:2015aha,Chen:2016zka}, and Higgs plus dijet production at NLO~\cite{Campbell:2006xx,vanDeurzen:2013rv} (for Higgs plus trijet production see \cite{Cullen:2013saa}). For inclusive Higgs production and for $\pTH < m_t$ (where $m_t$ is the top-quark mass) predictions in the full SM and in the HTL agree at the percent level. For $\pTH \ll m_t$ eventually fixed-order perturbation theory becomes unreliable and a matching to higher logarithmic accuracy becomes mandatory~\cite{Bozzi:2005wk,deFlorian:2011xf,Becher:2012yn,Bizon:2017rah,Chen:2018pzu,Bizon:2018foh,Re:2021con}, and also bottom-quark effects have to be considered~\cite{Mantler:2012bj,Grazzini:2013mca,Banfi:2013eda,Melnikov:2016qoc,Lindert:2017pky,Melnikov:2017pgf,Caola:2018zye}.    

At the other end of the spectrum, for $\pTH > m_t$, the accuracy of the HTL quickly deteriorates due to a different high-energy scaling compared to the full theory~\cite{Caola:2016upw}. For $\pTH=500(1000)\,\GeV$ the two differ by a factor of about 4(10). In this high-energy regime in order to improve with respect to the HTL additional $\mathcal{O}(1/m_t)$ corrections have been investigated ~\cite{Harlander:2012hf,Neumann:2014nha,Neumann:2016dny}.
Overall, the above cited explicit fixed-order computations have shown that
higher-order corrections computed in the HTL rescaled with lower-order
predictions with explicit mass dependence yield remarkably good approximations
of the full result despite the fact that the HTL is not valid in this energy regime. Therefore, it appears to be justified to tentatively apply this very same procedure also at the highest perturbative orders, where validation of the approximation is not yet possible. An example of such an approximation at the currently highest available perturbative order is presented in Ref. \cite{Becker:2020rjp} for Higgs plus jet production where NNLO corrections in the HTL are combined with NLO corrections in the full SM. However, defining a reliable uncertainty on such approximations remains crucial. Such approximations of reweighting higher-order computations in the HTL with exact lower order results are also at the basis of all currently available NLO Monte Carlo predictions matched to parton showers for Higgs plus (multi-)jet production~\cite{Campbell:2012am,Hamilton:2012rf,Buschmann:2014sia,Frederix:2016cnl,Greiner:2016awe}.

In this paper we present fixed-order NLO QCD computations for \pphj and also
\pphjj including top-quark mass effects. The computation for \pphj continues the
study of Ref.~\cite{Jones:2018hbb}, i.e. two-loop virtual corrections in the
full SM are evaluated numerically via
\secdec~\cite{Borowka:2015mxa,Borowka:2017idc}. Here we present additional
kinematic observables besides the \pTH distribution already shown in
Ref.~\cite{Jones:2018hbb} and compare the relative higher-order corrections in
the full SM against the HTL, and also an alternative approximation known as
\FTapprox, which has been introduced in Ref.~\cite{Maltoni:2014eza} in the
context of calculations for multi-Higgs production. In the \FTapprox all
ingredients of the NLO computation are computed exactly, except for the
two-loop virtual contributions, which are 
approximated in the HTL and reweighted with LO mass dependence. In the case of \pphjj production we compare results in the HTL and in the \FTapprox, as the exact five-point two-loop virtual amplitudes remain beyond current technology. The main aim of this study is three-fold: firstly, we would like to offer complementary kinematic information for the \pphj process, while also offering a (partial) validation of the results already presented in Ref.~\cite{Jones:2018hbb}. 
In this respect, the present study uncovered an issue affecting the real corrections included in Ref.~\cite{Jones:2018hbb}, which has subsequently been rectified. 
Secondly, we would like to investigate the \pphjj process in a kinematic regime relevant for VBF analyses. The comparison of NLO/LO ratios (usually known as K-factors) among HTL, \FTapprox and the full SM will help to put results obtained in a reweighted HTL on a more solid footing. Thirdly, results presented in this study can be seen as an intermediate step towards an NNLO computation of \pphj including exact mass effects wherever possible.

Technically, the computations of the NLO corrections to \pphj and \pphjj
production are performed within the \nnlojet fixed-order Monte Carlo framework,
which employs antenna subtraction for the handling of infrared (IR)
singularities~\cite{GehrmannDeRidder:2005cm,GehrmannDeRidder:2005aw,Daleo:2006xa,Daleo:2009yj,Glover:2010im,GehrmannDeRidder:2011aa,GehrmannDeRidder:2012ja,Ridder:2012dg,Currie:2013vh}.
All loop-squared amplitudes are evaluated via a new interface
between \nnlojet and
\openloops~\cite{Cascioli:2011va,Buccioni:2017yxi,Buccioni:2019sur} based on the latest (soon to be released) version \olotter
which in turn implements a new reduction method called \otter\cite{otter:2021}, which ensures excellent numerical stability in particular of the loop-induced real radiation amplitudes deep into the unresolved regime. We will investigate and discuss this numerical stability issue explicitly.

The structure of the paper is as follows. In Section~\ref{sec:tools} we describe the computational setup and employed tools. Numerical results for  $H$+jet and $H$+2~jets production will be presented in Section~\ref{sec:numerics}. We will conclude in Section~\ref{sec:conclusions}.


\section{Analysis framework and tools}
\label{sec:tools}

In this paper, we present predictions for the production of a boosted Higgs boson in the gluon-fusion channel.
We consider Higgs bosons produced in association with one (\pphj) or two (\pphjj) jets with NLO QCD corrections and include the effects of a finite top-quark mass either fully or via a suitable approximation.
For $pp \rightarrow H+j$ we compute the transverse momentum distribution of the Higgs boson including a finite top-quark mass, which has appeared previously in the literature~\cite{Jones:2018hbb,Lindert:2018iug,Neumann:2018bsx,Kerner:2019qgb}, as well as the Higgs boson plus jet invariant mass distribution.
For $pp \rightarrow H+jj$, the virtual corrections involve two-loop amplitudes for $2 \rightarrow 3$ scattering.
The mathematical complexity of the virtual corrections makes their computation currently intractable using either numerical or analytical methods.
We therefore adopt an approximation scheme to the full theory ($\mathrm{FT}_\mathrm{approx}$)~\cite{Frederix:2014hta,Maltoni:2014eza} for the NLO QCD corrections to Higgs boson plus two jet production.
Specifically, we include the exact top-quark mass dependence (SM) in the real corrections and infrared singular subtraction terms while using the virtual corrections in the heavy top-quark limit (HTL) re-weighted by the full Born level contribution on an event-by-event basis.
In fact, although the full matrix elements relevant to the virtual contributions of \hjj production are currently not available, nevertheless, their explicit infrared divergence at NLO can be predicted by the Catani dipole structure~\cite{Catani:1998bh}:
\begin{eqnarray}
	\text{Pole}\{|\mathcal{M}^{2}_{4}(m_t,\mu_R^2;\{p\})|^2\} = \sum \textbf{\emph{I}}^{(1)}(\epsilon,\mu_R^2;\{p\})|\mathcal{M}^{1}_{4}(m_t;\{p\})|^2,
\label{2loopSMpole}
\end{eqnarray}
where $m_t$ is the top-quark mass, $\mu_R^2$ is the renormalisation scale, $\{p\}$ is the momentum set regarding all external particles, $|\mathcal{M}^{m}_{n}|^2$ is the matrix element with $n$ legs and $m$ loops and $\textbf{\emph{I}}^{(1)}(\epsilon,\mu_R^2;\{p\})$ is the dipole operator containing all explicit IR divergences in $d$ space-time dimensions. The explicit expressions for dipole operators at squared matrix element level can be found in~\cite{GehrmannDeRidder:2005cm}.
We estimate the finite contribution of $|\mathcal{M}^{2}_{4}(m_t,\mu_R^2;\{p\})|^2$ by re-weighting the corresponding matrix element in the HTL approximation ($m_t\rightarrow \infty$) using:
\begin{eqnarray}
	|\mathcal{M}^{2}_{4}(m_t,\mu_R^2;\{p\})|^2 \rightarrow |\mathcal{M}^{1}_{4}(\infty,\mu_R^2;\{p\})|^2\frac{|\mathcal{M}^{1}_{4}(m_t;\{p\})|^2}{|\mathcal{M}^{0}_{4}(\infty;\{p\})|^2}.
\label{2loopHTLreweight}
\end{eqnarray}
Consequently, Eq.(\ref{2loopHTLreweight}) also recovers the explicit pole structure in Eq.(\ref{2loopSMpole}) and the explicit pole cancellation in the second bracket of Eq.(\ref{NLOsubstructure}) is automatically retained.

The $\mathrm{FT}_\mathrm{approx}$ scheme has proved to be remarkably reliable for Higgs plus one jet~\cite{Chen:2016zka} production and, to a lesser extent, di-Higgs~\cite{Grazzini:2018bsd} production (however in the latter case, it is much less reliable for differential distributions) at the LHC.
We implement and present the first application of this approximation to Higgs boson plus two jet production at NLO in QCD.


In the following sections we document the detailed implementation of our calculations. The \nnlojet program is used as a parton level event generator and all Born and real radiation one-loop contributions are computed using \olotter. The two-loop matrix elements involving a finite top-quark mass for \pphj are computed exactly using \secdec.

\subsection{Parton level event generator: \nnlojet}
\label{sec:nnlojet}
\nnlojet is a parton-level event generator equipped with flexible histogram analysis tools and scattering matrix elements evaluated in both the HTL approximation and in the full SM. 
It implements the antenna subtraction method to cancel infrared singularities from higher order QCD corrections~\cite{GehrmannDeRidder:2005cm,GehrmannDeRidder:2005aw,Daleo:2006xa,Daleo:2009yj,Glover:2010im,GehrmannDeRidder:2011aa,GehrmannDeRidder:2012ja,Ridder:2012dg,Currie:2013vh} while retaining the fully differential information of final state particles.

In this study, we combine \nnlojet with loop induced matrix elements provided by {\sc OpenLoops}2 and \secdec to study finite top-quark mass corrections for \hj and \hjj production with NLO QCD corrections.
The HTL and SM have the same infrared singular behaviour for both real emissions and virtual corrections, the antenna subtraction method can therefore be readily applied to regulate infrared singularities at NLO for loop-induced processes.  Schematically, the fully differential NLO contribution takes the form: 
\begin{eqnarray}
	\d\sigma_{H+n{\rm jet}}^{\NLO} & = & \int_{\d \Phi^{H}_{n+1}}\bigg[|\mathcal{M}^1_{n+3}(m_t;\{p\})|^2 - \sum_{\{p\}} X^0_3|\mathcal{M}^1_{n+2}(m_t;\{\tilde{p}\})|^2 \bigg]\nonumber\\
	                    & + & \int_{\d \Phi^{H}_{n}}\bigg[|\mathcal{M}^2_{n+2}(m_t,\mu_R^2;\{p\})|^2 + \sum_{\{p\}} \mathcal{X}^0_3|\mathcal{M}^1_{n+2}(m_t;\{p\})|^2\bigg],
\label{NLOsubstructure}
\end{eqnarray}
where $\Phi^{H}_{n}$ is the final state phase space of one Higgs plus $n$ partons, $X^0_3$ represents the tree-level three-parton antenna functions, $\mathcal{X}^0_3$ represents the corresponding integrated antenna functions in $d$-dimensions and $\{\tilde{p}\}$ is the momentum set after antenna mapping with one less parton compared to the $\{p\}$ momentum set. We use $X^0_3$ and the corresponding reduced matrix elements ($|\mathcal{M}^1_{n+2}|^2$) to capture the infrared singular behaviour of real radiations in $|\mathcal{M}^1_{n+3}|^2$, leading to a infrared finite contribution of the first bracket in Eq.(\ref{NLOsubstructure}). By adding back the integrated antenna functions $\mathcal{X}^0_3$ in the second bracket of Eq.(\ref{NLOsubstructure}), we render the integrand over the $\Phi^{H}_{n}$ phase space IR finite. The explicit IR divergences (which appear as poles in the regulator $\epsilon$) cancel analytically with the explicit IR poles from the virtual contribution ($|\mathcal{M}^2_{n+2}|^2$).

\subsection{One-loop contributions: \olotter}
All one-loop amplitudes contributing at the Born and real radiation level are
provided by the \olotter package, an upcoming improved version of the \openloops
program which implements a new reduction method called \otter \cite{otter:2021}.
Compared to the original
algorithm \cite{Cascioli:2011va}, {\sc OpenLoops}2  includes significant improvements
in numerical stability and
performance for the computation of
tree-loop interference amplitudes.
These improvements were achieved by a combination of the so-called on-the-fly reduction algorithm
\cite{Buccioni:2017yxi} and an automated stability system. 
However, the current implementation, which was designed for tree-loop interferences,
cannot be directly applied to loop-induced amplitudes, such as those required
for the present computation.
In \olotter a new tensor integral reduction method has been developed,
based on the on-the-fly reduction algorithm \cite{Buccioni:2017yxi}, that can
also be used for loop-squared amplitudes.  It profits from
recent improvements in performance and methods developed for handling numerical instabilities.
More specifically, numerical instabilities are avoided by
using certain freedoms in the selection of the reduction identities, and analytical any-order
expansions of three-point tensor integrals in the
limit of small Gram determinant \cite{Buccioni:2019sur}.
Furthermore, within \olotter residual instabilities are
captured by a rescaling test and tensor integrals are recomputed in
quadruple precision in case the given accuracy is not reached.
This upgrade to quadruple precision is efficient, and also important for
computations in deep infrared regions.  \olotter depends on {\sc Collier}
\cite{Denner:2016kdg} only for double precision scalar integrals and on {\sc OneLOop}
\cite{vanHameren:2010cp} for quadruple precision scalar integrals.

\paragraph{Numerical stability}

\begin{figure}[hbt!]
	\includegraphics[width=0.5\textwidth]{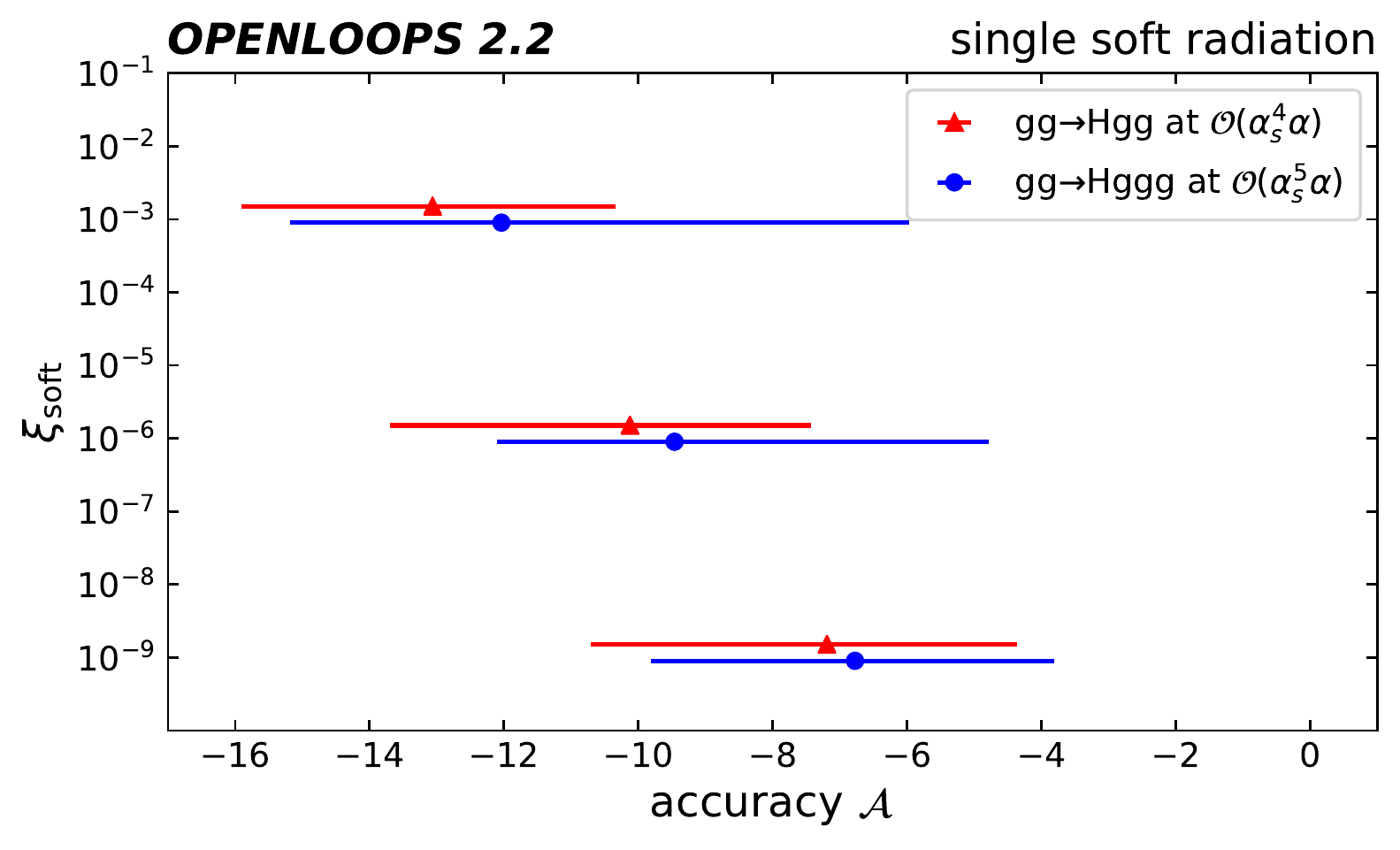}
	\includegraphics[width=0.5\textwidth]{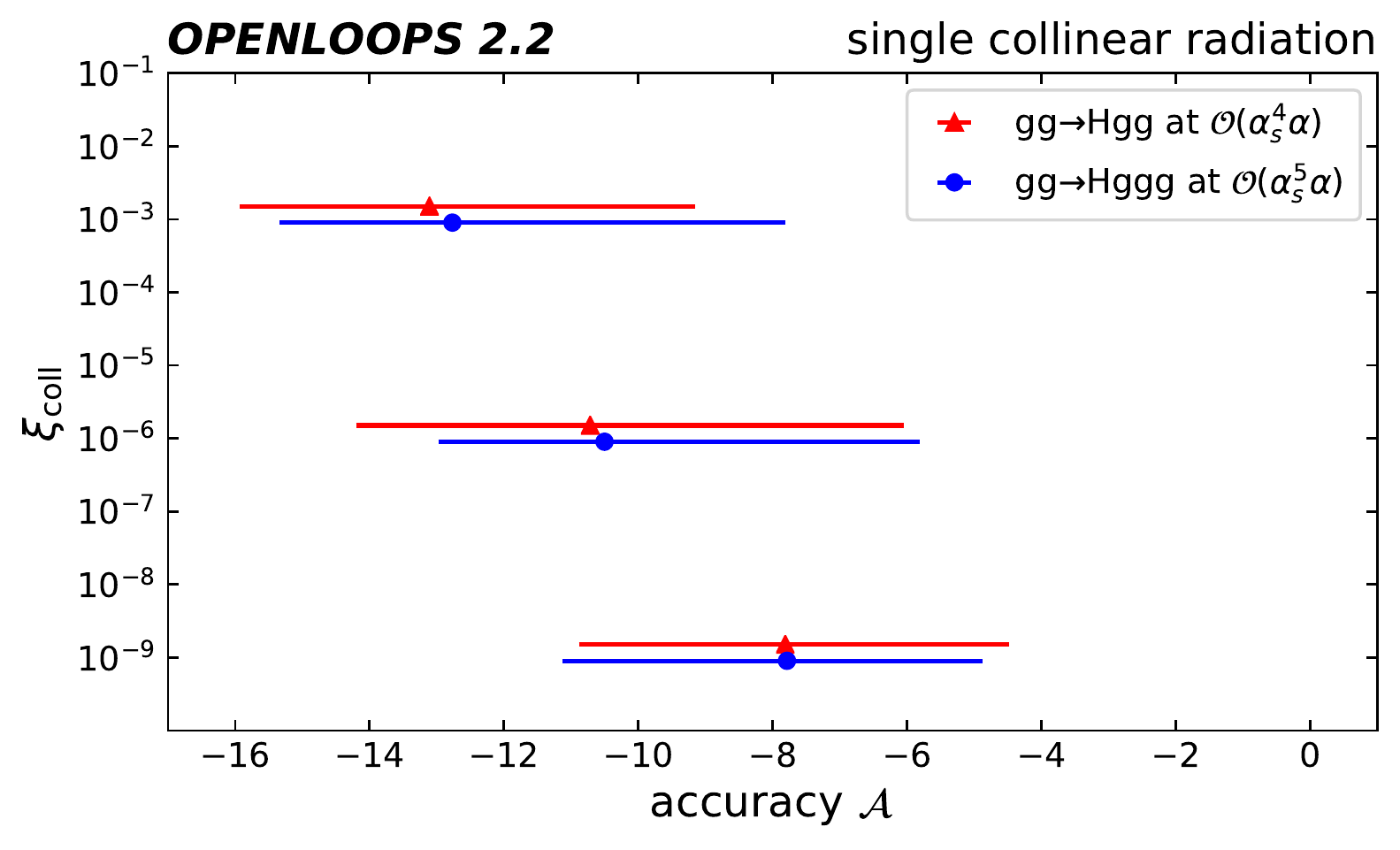}
\caption{Stability plots in IR regions for one-loop-squared matrix elements in
  $\mathrm{gg \to H gg}$ and $\mathrm{gg \to H ggg}$ versus the degree of
  collinear $\xi_{\text{coll}}$ or soft $\xi_{\text{soft}}$ singularity obtained
  with \olotter.
For each value of $\xi_{\text{soft/coll}}$, the numerical accuracy is calculated with a sample of $10^3$ randomly distributed infrared events. 
  Unstable points are detected by a rescaling test and rescued if the relative
  accuracy of $10^{-6}$ is not reached. The rescue step reevaluates 
  the tensor integrals to quad precision.
  The accuracy of the so-obtained value is determined by comparing
  it to a quadruple precision benchmark whose accuracy is also assessed by
  a rescaling test.
%
The plotted central points and variation bands correspond, respectively, to the average and 100\% confidence interval of $\mathcal{A}$.
}
\label{fig:IR_stability}
\end{figure}

The numerical stability of \olotter is crucial for the calculations of \hj and \hjj productions presented in this paper, especially to contributions from infrared kinematical regions, which are numerically challenging.
In Fig.~\ref{fig:IR_stability} we illustrate as a benchmark the stability 
of the critical $\mathrm{gg \to H gg}$ and $\mathrm{gg \to H ggg}$ amplitudes subject to single soft or collinear radiation. The degree of softness and collinearity are defined as
\begin{align}
  \xi_\mathrm{soft} = E_\mathrm{soft}/\sqrt{s}, \quad
  \xi_\mathrm{coll} = \theta_{ij}^2,
\end{align}
where $E_\mathrm{soft}$ is the energy of the soft particle, and $\theta_{ij}$ denotes the
angle of the collinear branching. 
The numerical stability is defined as 
\begin{align}
\mathcal A =  \log_{10} \Big( \frac{\mathcal W - \mathcal W^{(0)}}{\mathcal W^{(0)}} \Big),
\end{align}
where $\mathcal W$ denotes the one-loop-squared matrix element, $\mathcal W^{(0)}$ is the benchmark result, and $\mathcal A$ corresponds to the number of stable digits up to a minus sign. 
As can be seen from these plots, the numerical accuracy remains very high all the way down to the deep infrared regime. 

\paragraph{Numerical performance}

In Table~\ref{tableperf} we present results for the average evaluation time of samples of random phase-space points using various different modes/versions of {\sc OpenLoops}.
In particular, here certain parts of the amplitude are evaluated in double or in quadruple
precision, or a realistic error estimate of the amplitudes is performed.
In summary, performance is greatly improved in \olotter which
in particular makes a tensor integral based rescaling test cheap.
Moreover, since \olotter operates also in quadruple precision with very high
numerical efficiency, numerically unstable points can be rescued in a reliable
way, which has largely been prohibitive for loop-squared amplitudes in {\sc
OpenLoops2}.
In fact, \olotter allows for rescue of unstable points in
a new hybrid mode, where only the tensor integrals are evaluated in higher
numerical precision resulting in a 8-fold and 3-fold increase in runtime
compared to pure double precision for $gg \rightarrow Hgg$ and $gg \rightarrow
Hggg$ respectively, compared to a roughly 80-fold increase in runtime for full
quadruple precision.
In practice, and as used for the present computation of this
paper in \olotter a combination of pure double precision with this new hybrid mode is used. Based on a pure double precision evaluation the stability for every phase-space point is estimated based on a rescaling test of only the tensor integrals. 
Then, only for critical points the tensor integrals are reevaluated in quadruple precision based on the hybrid mode.
\begin{table}[h!]
\centering
  \begin{tabular}{  lccc | ccc }
    Mode & $gg \rightarrow Hgg$ (time/psp) & $gg \rightarrow Hggg$ (time/psp)\\ \hline\hline
    OL2.1+Collier DP                    & 13ms & 0.56s \\
    OL2.1+Collier DP + error estimation & 19ms & 0.89s \\
    OL2.1+CutTools QP                   & 43000ms  & 2300s \\
\hline
    OL2.2+Otter DP                      & 8.9ms  & 0.29s \\
    OL2.2+Otter DP + error estimation   & 11ms & 0.32s \\
    OL2.2+Otter DP+QP tensor integrals  & 68ms & 0.87s \\
    OL2.2+Otter QP                      & 740ms  & 23s \\
  \end{tabular}
   \caption{Runtimes for loop-squared amplitudes for $g g\to Hgg$ and $g g\to Hg
   gg$ in {\sc OpenLoops}.
  All numbers have been produced on an Intel(R) Core(TM) i7-7700 CPU @ 3.60GHz.
  The first three rows correspond to the modes so far available in {\sc
  OpenLoops}2~\cite{Buccioni:2019sur}. These employ double precision evaluation
  with {\sc Collier} (first and second row), where in the second row tensor
  integrals are computed twice using the COLI and DD branches of {\sc Collier}
  in order to obtain an error estimate. The third row employs {\sc CutTools} and
  the entire amplitude is evaluated in quadruple precision. For the amplitudes
  at hand the resulting runtimes in quadruple precision are prohibitive to be
  used as rescue system. The lower four rows represent evaluation based on the
  new \otter method in \olotter. In this case the error estimation is performed via a rescaling test where all tensor integrals are recomputed with rescaled kinematics.  The sixth row corresponds to a new hybrid mode where only the tensor integrals are evaluated in quadruple precision, and everything else in double precision.
  The last row shows the performance for a full quadruple precision
  evaluation within \olotter.
   }
   \label{tableperf}
\end{table}


\subsection{Two-loop contributions: \secdec}
\label{sec:secdec}

For \pphj production, we evaluate the two-loop virtual contributions with exact top-quark mass dependence as presented in Refs.~\cite{Jones:2018hbb,Kerner:2019qgb}.
Briefly, the two-loop amplitudes, which depend on four mass scales (the Mandelstam invariants $s$ and $t$ as well as the two masses $m_t$ and $m_h$), are expressed in terms of a basis of master integrals using the program {\sc Reduze}2~\cite{vonManteuffel:2012np}.
In order to obtain the integral reduction in a reasonable time and to reduce the size of the resulting amplitude, the ratio of the Higgs boson mass to the top-quark mass is fixed according to $m_H^2/m_t^2 = 12/23$.
The master integrals are then sector decomposed using the program \secdec~\cite{Borowka:2015mxa,Borowka:2017idc} and numerically integrated on Graphics Processing Unit (GPUs) using the Quasi-Monte Carlo method~\cite{Li:2015foa,Borowka:2018goh}.
To improve the stability of the amplitude we select a quasi-finite basis of master integrals as outlined in Ref.~\cite{Kerner:2019qgb}, this differs from the basis originally used in Ref.~\cite{Jones:2018hbb}.
We observe that the new choice of master integrals also significantly reduces the complexity of the coefficients of the master integrals appearing in the amplitude and thus the size of the code.

The results presented here are produced using a total of 6497 phase-space points for the two-loop virtual contribution.
In Ref~\cite{Jones:2018hbb}, a fraction of the phase-space points were distributed such that they provide a good estimate of the total cross section (assuming a jet cut of $p_{\rm{T},j}>30\,\GeV$) and additional phase-space points were generated to sample the tail of the $p_{\rm{T},H}$ distribution.
We reuse these existing phase-space points and also compute an additional 1007 points to populate the large invariant mass region $m_{Hj}$ for $p_{\rm{T},j} > 300\,\GeV$.

\section{Numerical results}
\label{sec:numerics}

\subsection{Setup}
\label{sec:setup}
As an extension of the study of Higgs plus one jet production at NLO~\cite{Jones:2018hbb}, we adopt the same input parameters and numerical setup in the current calculation. 
To quantify the impact of increasing the number of final state jets, we keep the input parameters consistent between \hj and \hjj production. 
The counting of the number of jets in this study is inclusive. There is no difference between inclusive and exclusive jet counting for LO while results at NLO accuracy receive contributions from real emissions including events classified with one addtional jet.
For the Higgs and top-quark mass  we use $m_H=125\,\GeV$ and $m_t={173.055}\,\GeV$.
The top-quark Yukawa coupling $\lambda_t=\sqrt{2} m_t/v$ is determined by the vacuum expectation value (vev) of the Higgs $v=\frac{ M_W \sin\theta_W}{\sqrt{\pi} \alpha_{\rm QED}}=246.219\,$GeV and the top-quark mass. We use the five flavour scheme assuming light quarks are massless in both inital and final states.


Throughout our calculation, the top-quark mass is renormalised using the on-shell (or pole mass) scheme. 
It has been pointed out in the literature that several Higgs boson production processes (including Higgs boson production in association with jets) are sensitive to the scheme and scale used to renormalise quark masses~\cite{Baglio:2018lrj,Baglio:2020ini,Baglio:2020wgt,Amoroso:2020lgh}. 
For example, at LO, the difference between the pole mass scheme and the $\overline{\mathrm{MS}}$ scheme at scale $m_{H,j}/2$ was found to be around $12\%$ for $m_{H,j}=700\,\GeV$ and $p_{T,j} > 300\,\GeV$. 
At larger $p_{T,H}$ the difference between the two schemes grows and can reach $\approx 25\%$ for $p_{T,H}=1\,\TeV$. 
In off-shell Higgs boson production, off-shell Higgs boson decay to photons and in Higgs pair production, the NLO corrections reduce the mass scheme uncertainty to approximately half that of the LO~\cite{Baglio:2018lrj,Amoroso:2020lgh}. 
By analogy, we may expect that the mass scheme uncertainty, which we do not assess, is similar in size to our NLO scale uncertainties (see below).

We employ the PDF4LHC15\_nlo\_30\_pdfas PDF set~\cite{Butterworth:2015oua} throughout 
and all of our predictions are at a center of mass energy of $\sqrt{S}=13$\,TeV. Renormalisation and factorisation scales are chosen as
\begin{align}
&\mu_{\rm R,F}=\xi_{\rm R,F} \cdot H_T/2, \quad \mathrm{with} \quad H_T =	\sqrt{m_H^2+p_{\rm{T},H}^2} + \sum\limits_{j} |p_{\rm{T},j}|\,,
\label{eq:centralscale}
\end{align}
where the sum includes all final state partons. Our central scale corresponds to $\xi_{\rm R,F}=(1,1)$ and we determine scale uncertainties via the standard 7-point factor-2 variations $\xi_{\rm R,F}=(2, 2), (2, 1), (1, 2),$ $ (1, 1), (1,\frac{1}{2}),$ $(\frac{1}{2}, 1),
(\frac{1}{2},\frac{1}{2}))$.
Any reconstructed jets are clustered via the anti-k$_t$~\cite{Cacciari:2008gp} algorithm with $R=0.4$. We apply the following cuts:
\begin{align}
	 H+\mathrm{jet}: & \quad p_{\rm{T},j}>30\,\GeV,\\
	 H+2\,\mathrm{jets}: & \quad p_{\rm{T},j_1}>40\,\GeV \  \mathrm{and} \  p_{\rm{T},j_2}>30\,\GeV\,.
\end{align}
The latter asymmetric jet cuts avoid a perturbative instability in the limit 
$p_{\rm{T},H} \to 0$\,\GeV.

\subsection{Fiducial total cross sections}

\renewcommand{\arraystretch}{1.7}
\begin{table}[t]
\centering
  \begin{tabular}{  ll|ccc | ccc }
    \multirow{2}{*}{\large$\sigma$[pb]} && \multicolumn{3}{c|}{Inclusive $p_{T,H}$} & \multicolumn{3}{c}{$p_{T,H}>300$ GeV}\\ 
    
              &   & LO & NLO & K & LO & NLO & K\\ \hline\hline
              
    \multirow{3}{*}{\hj} &HTL & $8.22^{+3.17}_{-2.15}$ & $13.57^{+2.11}_{-2.09}$  & $1.65$  & $0.086^{+0.038}_{-0.024}$ & $0.160^{+0.033}_{-0.030}$ & $1.86$  \\ 
     &\FTapprox & $8.56^{+3.30}_{-2.24}$ & $14.06(1)^{+2.17}_{-2.16}$ & $1.64$ & $0.046^{+0.020}_{-0.013}$ & $0.088^{+0.019}_{-0.017}$ & $1.91$ \\
     & SM & $8.56^{+3.30}_{-2.24}$ &  $14.15(7)^{+2.29}_{-2.21}$ & $1.65$ & $0.046^{+0.020}_{-0.013}$ & $0.089(3)^{+0.020}_{-0.017}$ & $1.93$ \\    \hline
        \multirow{3}{*}{\hjj} &HTL & $2.87^{+1.67}_{-0.99}$ & $4.33^{+0.59}_{-0.80}$ & $1.51$ & $0.120^{+0.071}_{-0.042}$  & $0.160^{+0.012}_{-0.025}$ & $1.33$ \\
     &\FTapprox & $2.92^{+1.70}_{-1.01}$ & $4.45(1)^{+0.63}_{-0.83}$ & $1.52$ & $0.068^{+0.040}_{-0.024}$ & $0.092^{+0.008}_{-0.015}$ & $1.35$ \\
     & SM & $2.92^{+1.70}_{-1.01}$ &  $-$ & $-$ & $0.068^{+0.040}_{-0.024}$ & $-$ & $-$   \end{tabular}
   \caption{\label{tab:totalXS}Integrated cross sections at LO and NLO in the HTL and \FTapprox approximations and with full top-quark mass dependence (SM) for \hj and \hjj production together with corresponding K-factors. Uncertainties correspond to the envelope of 7-point scale variations. For \hj production we require $p_{\rm{T},j}>30\,\GeV$, while for \hjj production we require $p_{\rm{T},j_1}>40\,\GeV,\, p_{\rm{T},j_2}>30\,\GeV$. On the left no further phase-space restrictions are considered, while on the right we additionally require $p_{T,H}>300\,\GeV$. Numerical integration errors larger than permil level are indicated in brackets.
   }
\end{table}

Applying the computational setup in Section~\ref{sec:setup}, we document the fiducial total cross section for \hj and \hjj produciton in Tab.~\ref{tab:totalXS}. For reference we present results for the HTL, \FTapprox and SM predictions. We also include the fiducial total cross section for boosted Higgs with $p_{T,H}>300$~GeV. Further fiducial cross sections with varying $p_{T,H}$ cuts are listed in Appendix~\ref{app:xsections}. 
The numbers for \hj reported here agree, within the statistical uncertainty, with the updated version of Ref.~\cite{Jones:2018hbb}. 

Together with predictions obtained with the central scale defined in Eq.~\eqref{eq:centralscale} we show the upper and lower values obtained by the envelope of 7-point scale variations. 
For \hj production without the fiducial constraint of $p_{T,H}$, the top-quark mass effects lead to an increase of 4.3\% (4.6\%) at LO (NLO) comparing to HTL. There is an increase of about 1\% in the total NLO cross section when comparing the \FTapprox result with the full top-quark mass dependence. The NLO/LO K-factor is consistent among HTL, \FTapprox and SM at about 1.65.

With the fiducial constraint of $p_{T,H}$ larger than 300 GeV, we observe a similar amount (2.3\%) of relative increase from the \FTapprox prediction to the result with full top-quark mass effects at NLO. In contrast, the HTL prediction is 78\% (81.8\%) larger than the full theory (\FTapprox) result due to large logarithmic corrections from the disparity in scales between $m_t^2$ and $p_{T,H}^2$. The NLO/LO K-factor for $p_{T,H}> 300$ GeV is however again almost universal among HTL, \FTapprox and SM at about 1.9. To be precise, the K-factor in the full SM is $5.3\%$ resp. $2.6\%$ larger compared to the HTL and \FTapprox predictions.     

For \hjj production without the fiducial constraint on $p_{T,H}$, the \FTapprox prediction induces a 1.7\% (2.8\%) increase compared the HTL approximation at LO (NLO). Considering the fiducial constraint of $p_{T,H}$ larger than 300 GeV, the HTL fiducial total cross section is about 1.76 (1.74) times the \FTapprox predictions at LO (NLO). However, again in both selections NLO K-factors are universal with mass corrections in the \FTapprox below the $2\%$ level.

For the absolute cross sections the extra jet emission in \hjj production decreases the total \hj cross section by more than a factor of 3 from $14.06$ pb (\hj) to $4.45$ pb considering the \FTapprox predictions at NLO in both cases.
In contrast, in the boosted Higgs boson regime, the total NLO cross sections are comparable, while the NLO K-factor is about $46\%$ larger in the \hj computation compared to the \hjj one. This can be understood from the restriction to the back-to-back configuration in the \hj computation at LO. Multi-jet configurations recoiling against the hard Higgs only open up at NLO, where they are effectively described at LO. In contrast in the \hjj computation such configurations already contribute at LO.


In order to quantify top-quark mass effects in various fiducial regions and to explore the possibility to extrapolate the mass effect to higher order corrections, in the following sections we present in-depth comparisons of differential cross sections for \hj and \hjj production.

\subsection{Fiducial differential cross sections for \hj production}
\label{sec:hjdiff}

\begin{figure}[hbt!]
	\includegraphics[width=0.5\textwidth]{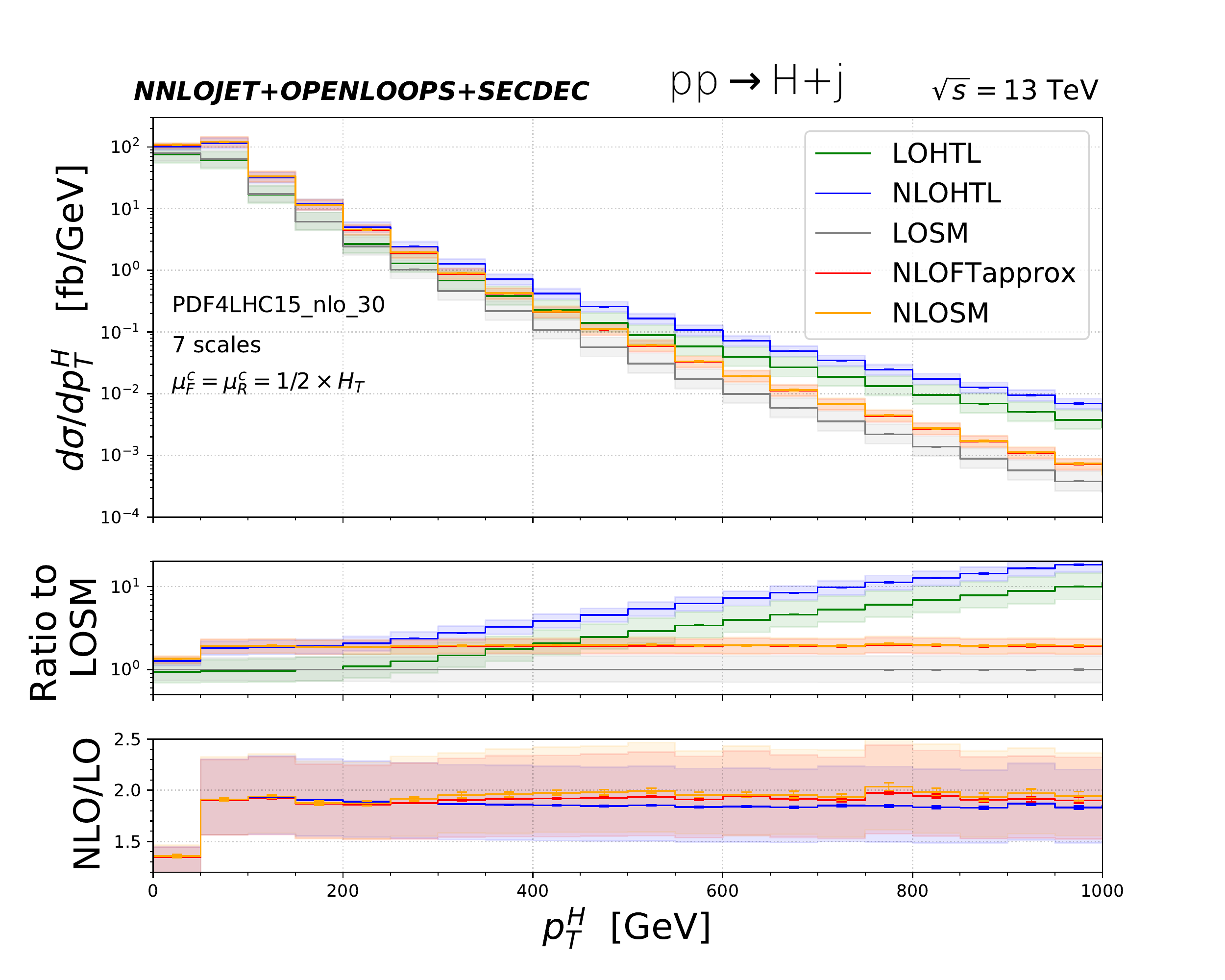}
\includegraphics[width=0.5\textwidth]{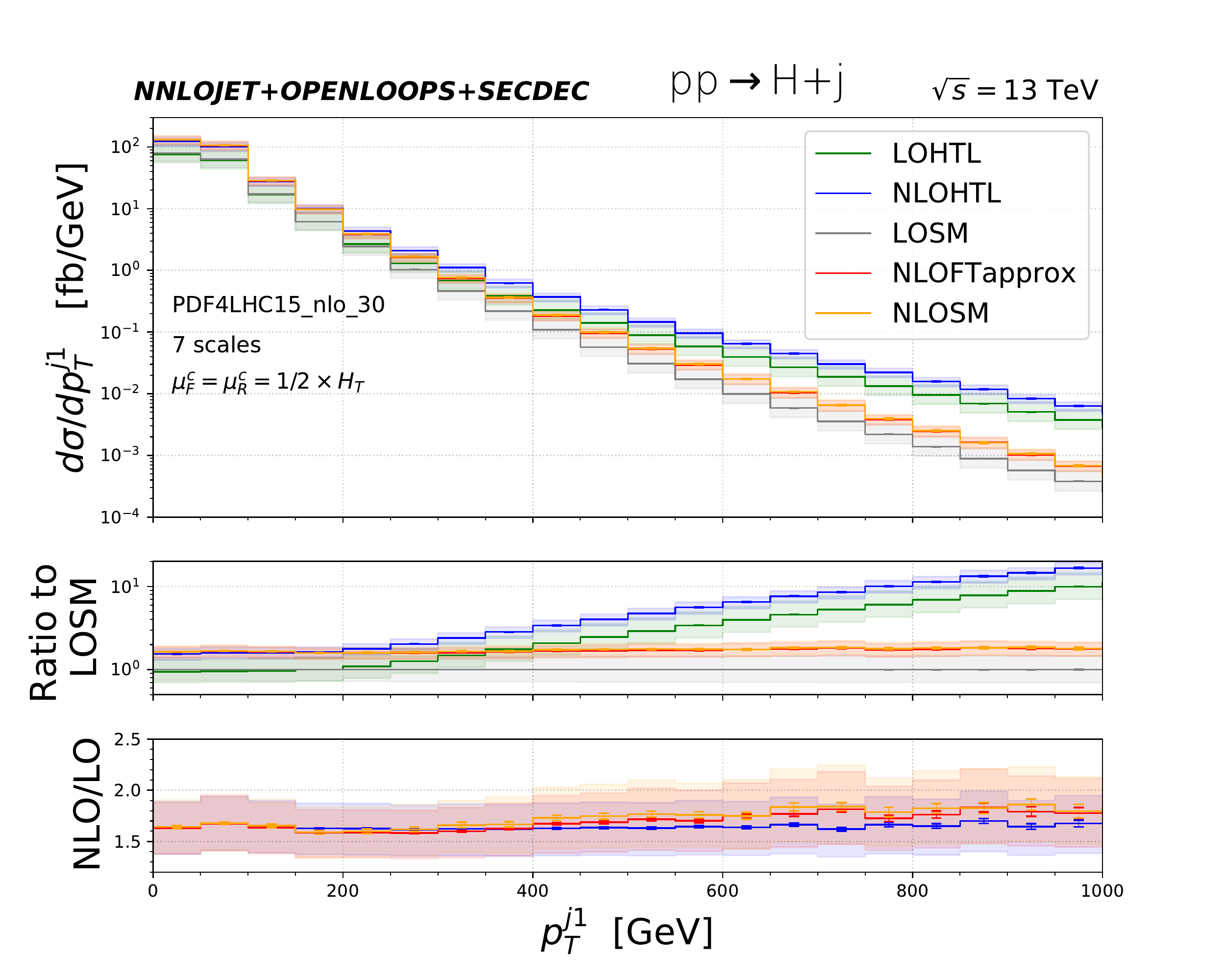}
\caption{Transverse momentum distribution of the Higgs (left) and the hardest jet (right) in \hj production. We show LO predictions in the full SM (LOSM, magenta) and the HTL (LOHTL, green) as well as NLO predictions in the \HTL (NLOHTL, blue), the \FTapprox (NLOFTapprox, red) and the full SM (NLOSM, orange). The upper panel shows absolute predictions. The first ratio plot shows corrections with respect to LOSM, while the second ratio plot shows NLO corrections normalised to the respective LO prediction, i.e. NLOHTL/LOHTL, NLOSM/LOSM, and NLOFTapprox/LOSM. Shaded bands correspond to scale variations. Error bars indicate integration uncertainties.}
\label{fig:hj_pth_ptj}
\end{figure}

\begin{figure}[hbt!]
	\includegraphics[width=0.5\textwidth]{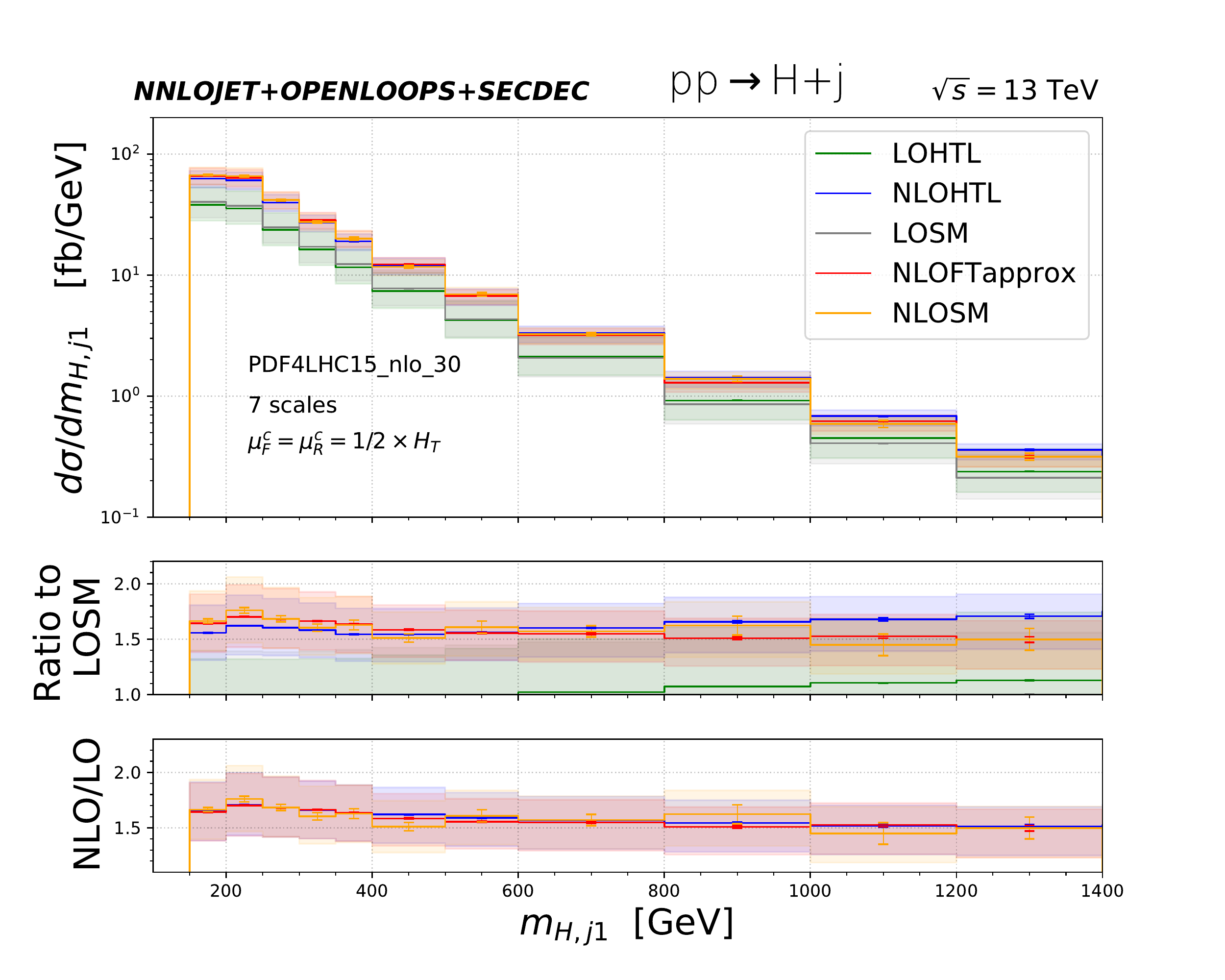}
		\includegraphics[width=0.5\textwidth]{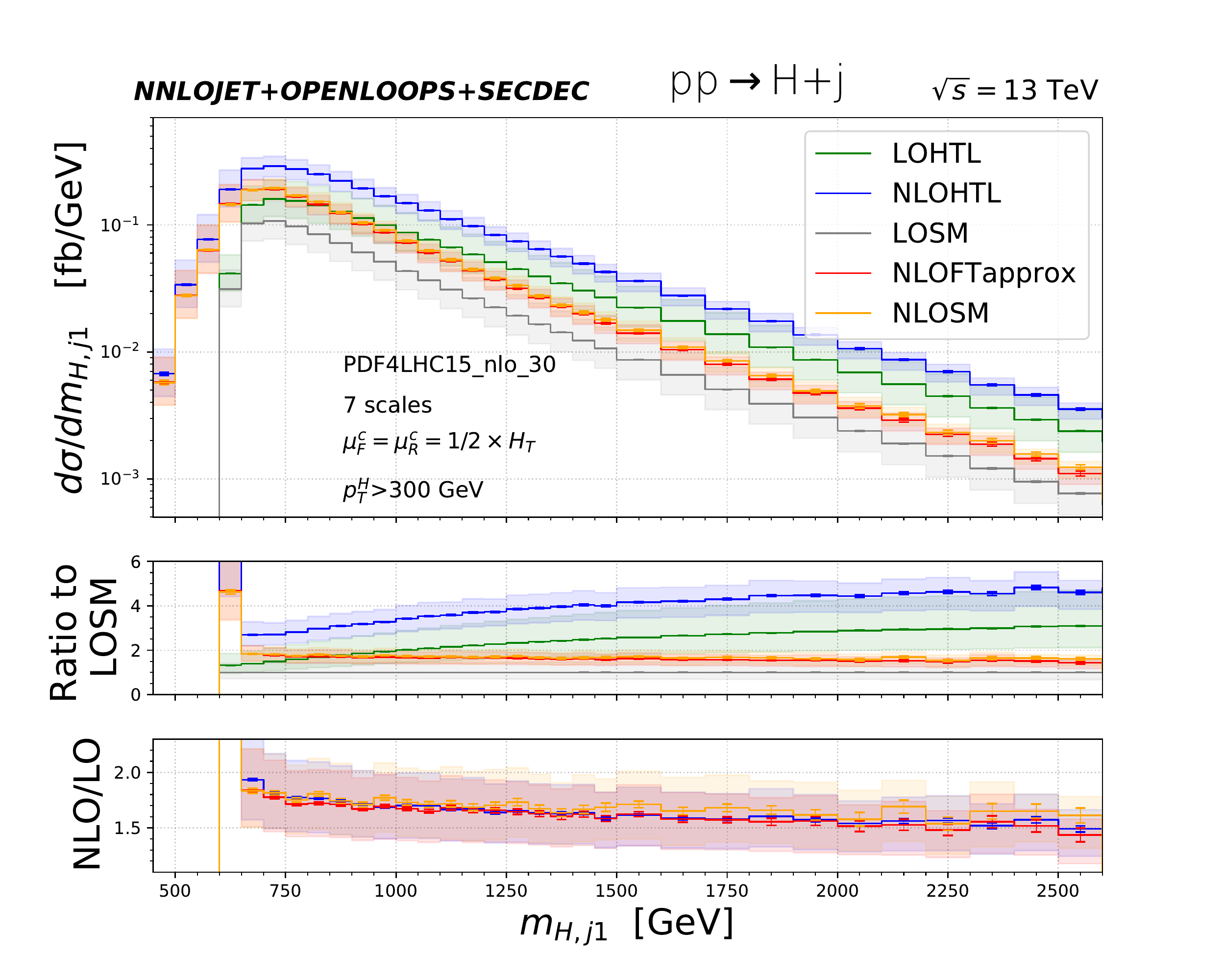}
\caption{Invariant mass distribution of the Higgs and the jet for inclusive  \hj production (left) and with $\pTH > 300\,\GeV$ (right). Colour coding and labelling  as in Fig.~\ref{fig:hj_pth_ptj}.}
\label{fig:hj_mj}
\end{figure}

In Figs.~\ref{fig:hj_pth_ptj}-\ref{fig:hj_mj} we present numerical results for \hj production at the LHC with $\sqrt{S}=13$\,TeV. We compare NLO corrections in the \HTL with the \FTapprox approximation, and with the full SM. 

In Fig.~\ref{fig:hj_pth_ptj} we consider the transverse momentum distribution of the Higgs on the left and of the jet on the right. These distributions are identical at LO, and are also highly correlated at NLO. As is well known, for $p_{T,X} > m_t\ (X=H,j)$ the HTL approximation breaks down and predicts a very different high-energy scaling compared to the SM. However, for NLO/LO ratios in the bottom panels of Fig.~\ref{fig:hj_pth_ptj}, at least at the $\ord(10\%)$ level, higher-order QCD corrections in the \HTL agree well with the \FTapprox and the full SM. Further improvement of agreement is observed between the \FTapprox and the full SM at the $\ord(5\%)$ level within numerial uncertainties. The scale variation bands in the ratio plots are obtained by fixing the observable at the central scale choice in the denominator while taking the envelope of scale choices in the numerator by the 7-point factor-2 variations. We observe consistent agreement for the size of scale variations also for the ratio plots. In \pTH the corrections are at the level of $90-110\%$, while in \pTj they are at the level of $60-80\%$, for the entire considered range, i.e shape corrections are mild. Examining the corrections in more detail an overall increase of the corrections of $\approx 5-10\%$ can be appreciated for the full SM compared to the HTL, with a mild relative increase at large transverse momenta slightly larger for \pTj than for \pTH. Corrections in the \FTapprox and the HTL agree exactly up to $\pTH/\pTj \sim m_t$. Beyond that, corrections in the \FTapprox are a few percent larger compared to the HTL, with again a very slight increase at large transverse momenta. 
The remaining scale uncertainties at NLO are at the $20-25\%$ level throughout. 

In Fig.~\ref{fig:hj_mj} we turn to a different kinematic regime in \hj production by considering invariant mass distributions in the Higgs-jet system on the left for inclusive \hj production and on the right considering \pTHhigh. For the inclusive \mHj distribution the HTL NLO result increases 
mildly by $10-20\%$ in the tail of the distribution compared to the \FTapprox and the full SM. However, the NLO K-factor is universal in all three predictions decreasing from about $1.7$ at small \mHj to about $1.5$ at large \mHj with similar relative size of scale variation.
Finally, in the exclusive \pTHhigh phase-space absolute predictions in the HTL and \FTapprox diverge for increasing \mHj. 
However, also here the relative NLO corrections are found to be largely identical for the HTL, \FTapprox and full top-quark mass results. This holds at large \mHj as well as at small \mHj, which for $\mHj < 600$ GeV is kinematically inaccessible at LO due to the \pTHhigh requirement. 

From the detailed comparison of \pTH, \pTj and \mHj distributions of \hj production, we observe excellent agreement of differential NLO/LO K-factors (for the central scale and scale variations) among theory predictions using HTL, \FTapprox and the exact top-quark mass dependence. This observation validates the multiplicative reweighting procedure introduced in~\cite{Chen:2016zka} at histogram level and further strengthens the reweighed predictions for \hj production at NNLO accuracy~\cite{Becker:2020rjp}.

%
%


\subsection{Fiducial differential cross sections for \hjj production}
\label{sec:hjjdiff}
\begin{figure}[hbt!]
	\includegraphics[width=0.5\textwidth]{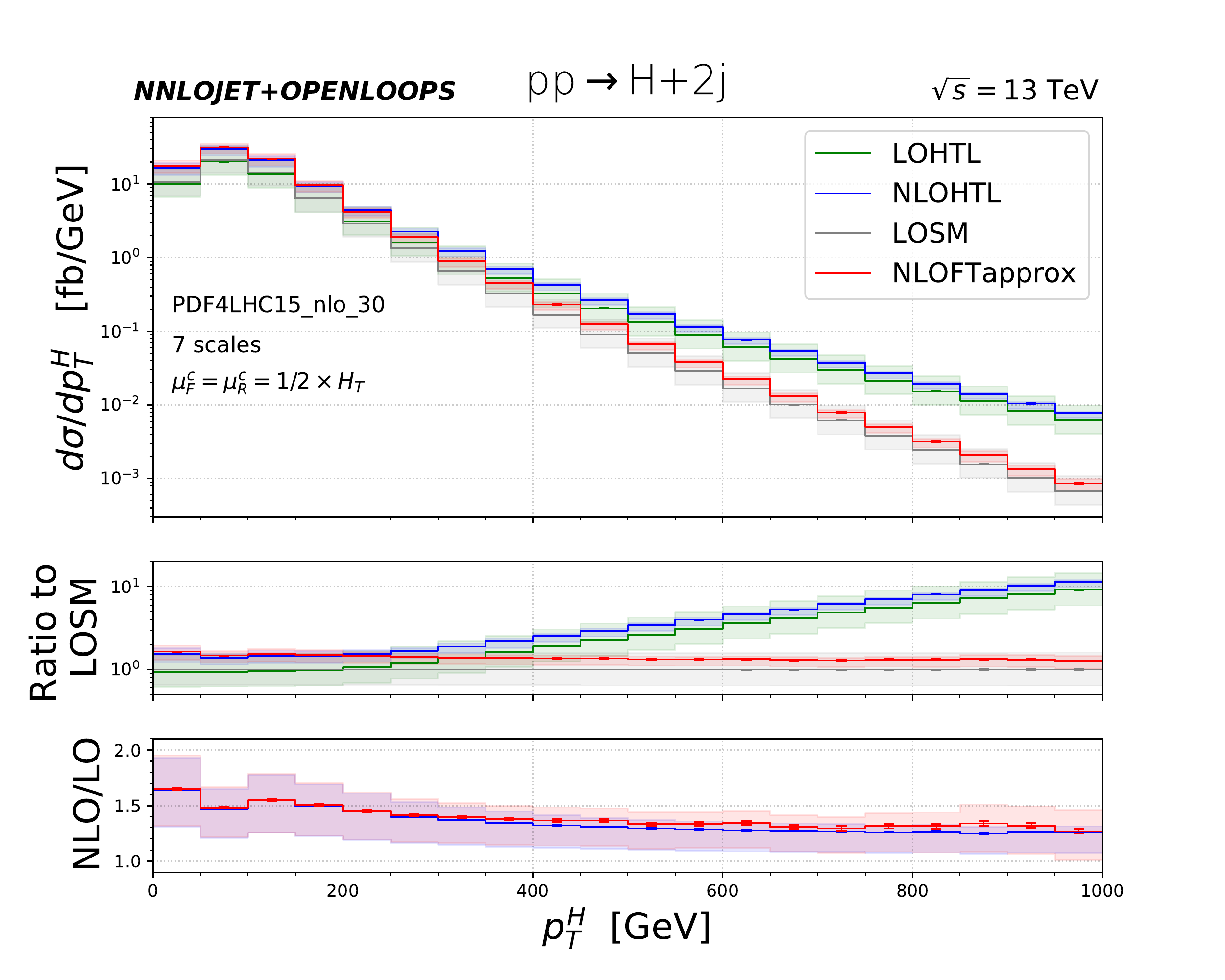}
\includegraphics[width=0.5\textwidth]{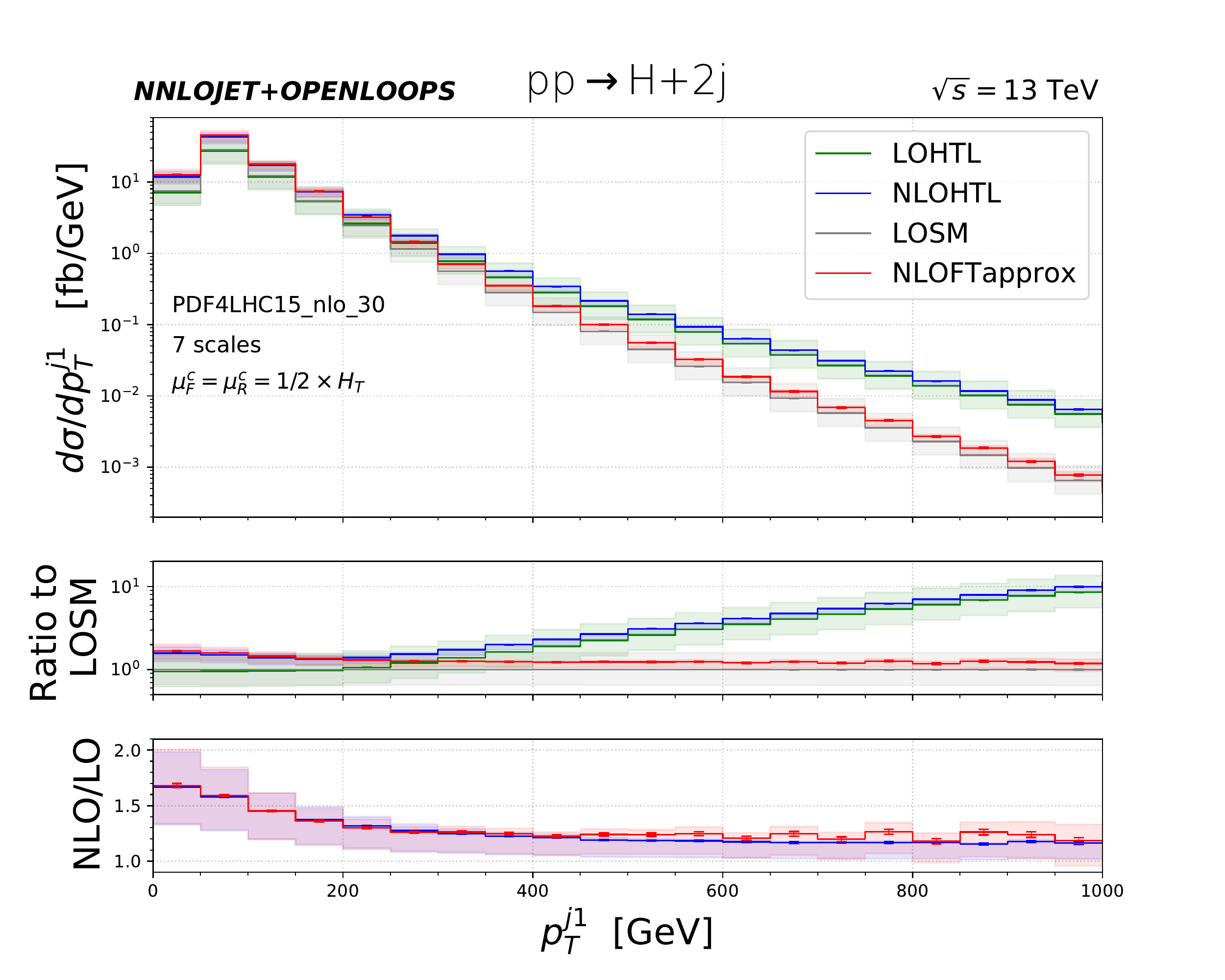}
\caption{Transverse momentum distribution of the Higgs (left) and the hardest jet (right) in \hjj production. Colour coding and labelling  as in Fig.~\ref{fig:hj_pth_ptj}.}
\label{fig:hjj_pth_ptj}
\end{figure}

\begin{figure}[hbt!]
	\includegraphics[width=0.5\textwidth]{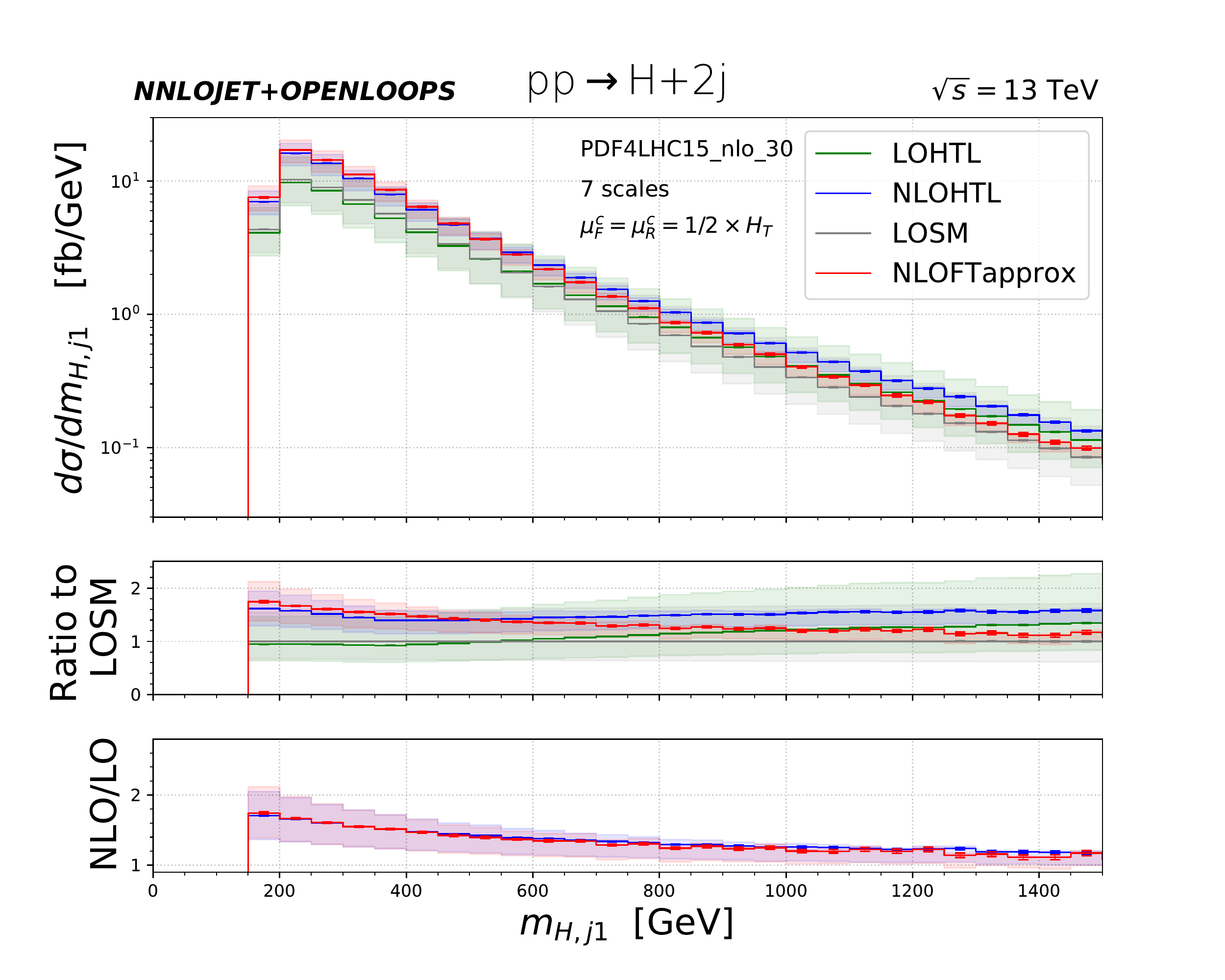}
		\includegraphics[width=0.5\textwidth]{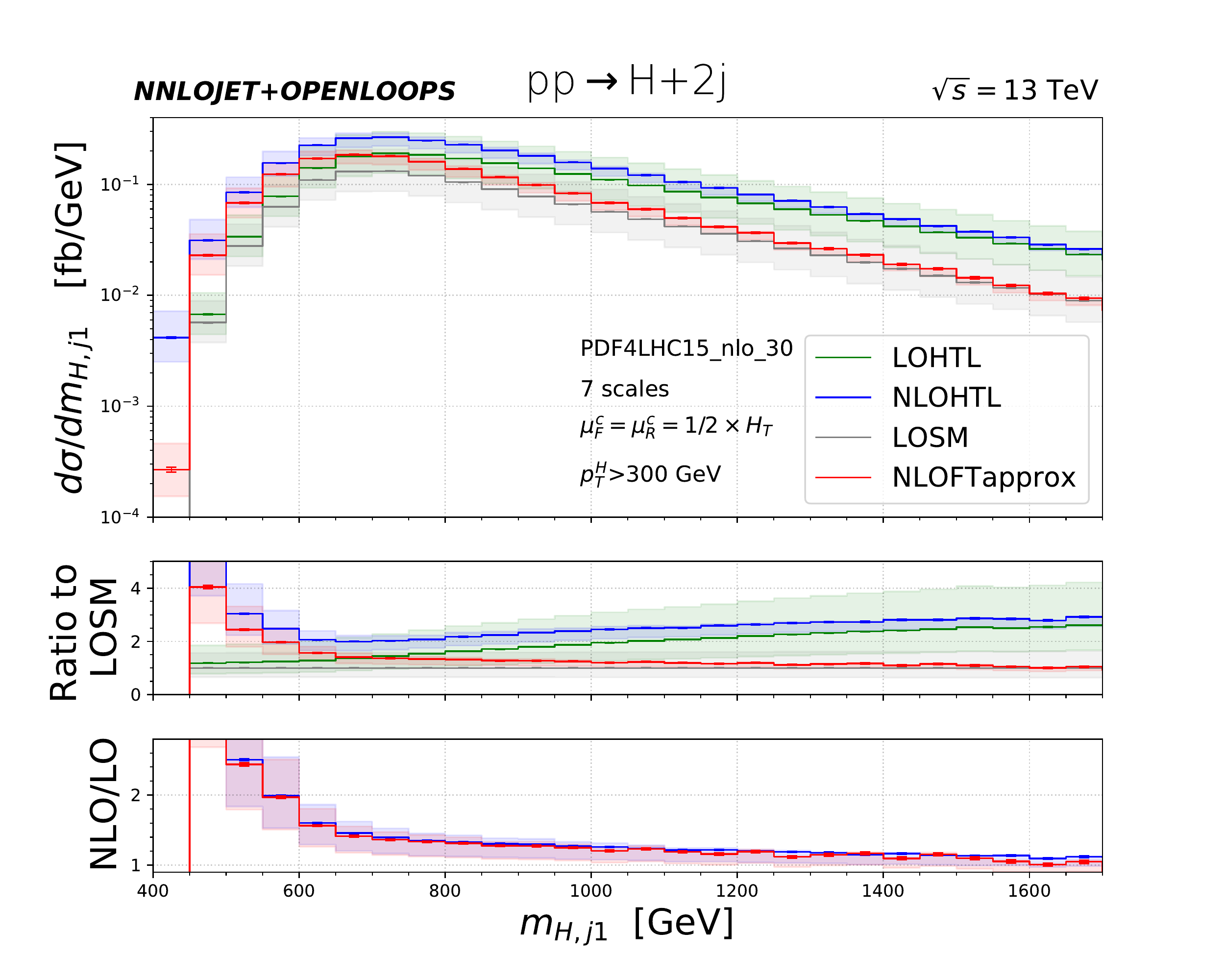}
\caption{Invariant mass distribution of the Higgs and the hardest jet system in \hjj production. Colour coding and labelling as in Fig.~\ref{fig:hj_pth_ptj}.}
\label{fig:hjj_mj}
\end{figure}

\begin{figure}[hbt!]
	\includegraphics[width=0.5\textwidth]{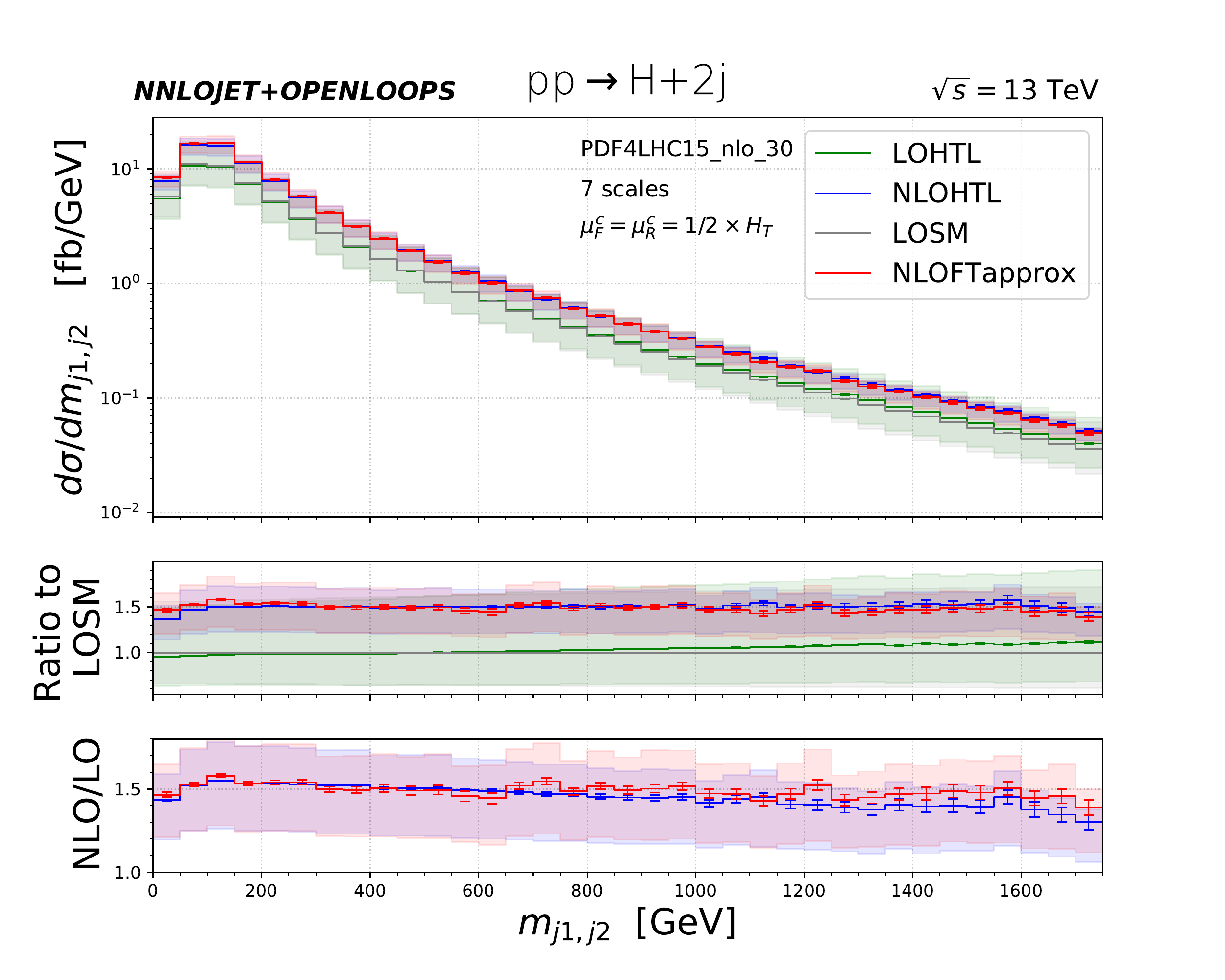}
		\includegraphics[width=0.5\textwidth]{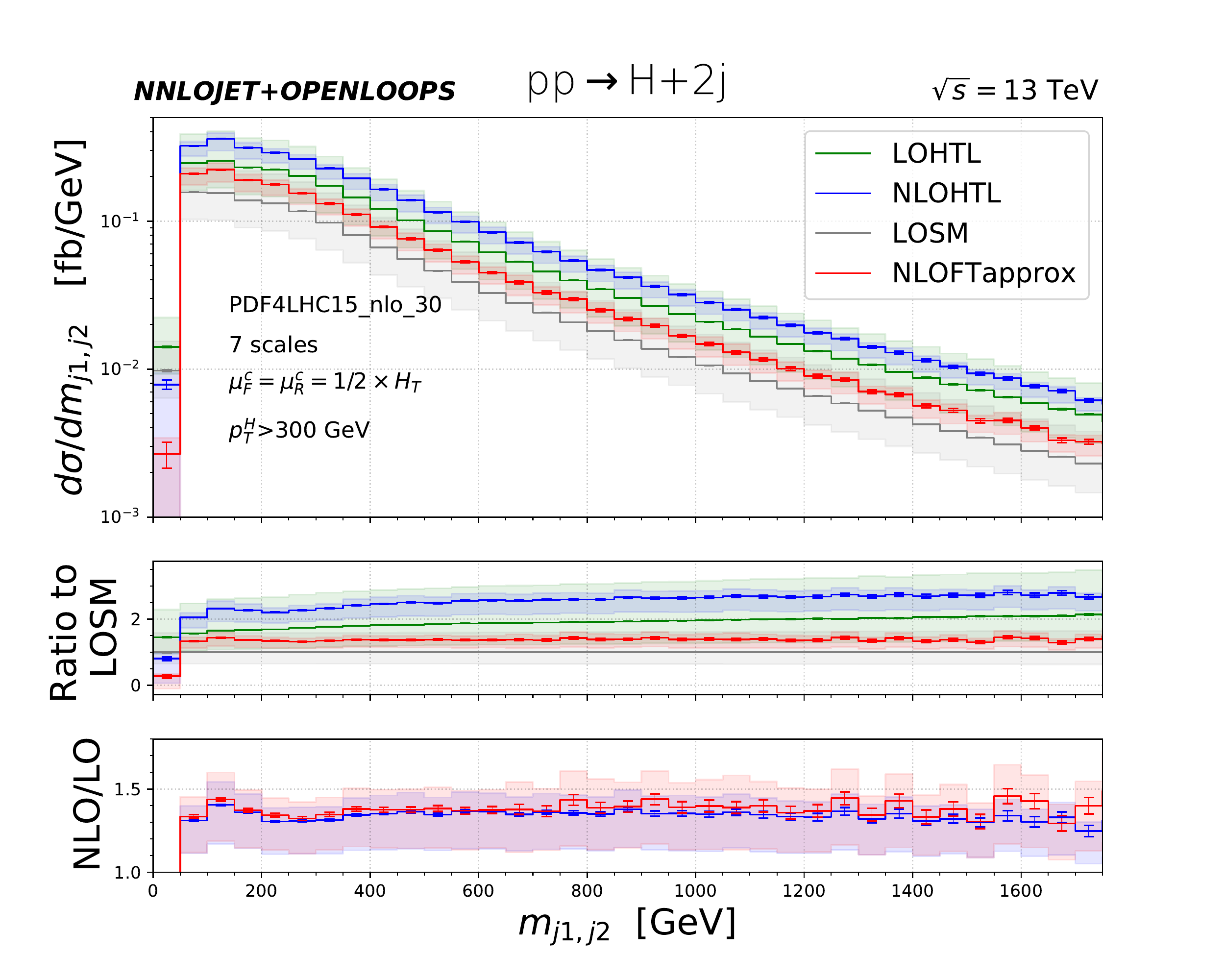}
\caption{Invariant mass distribution of the di-jet system in \hjj production. Colour coding and labelling as in Fig.~\ref{fig:hj_pth_ptj}.}
\label{fig:hjj_mjj}
\end{figure}

\begin{figure}[hbt!]
	\includegraphics[width=0.5\textwidth]{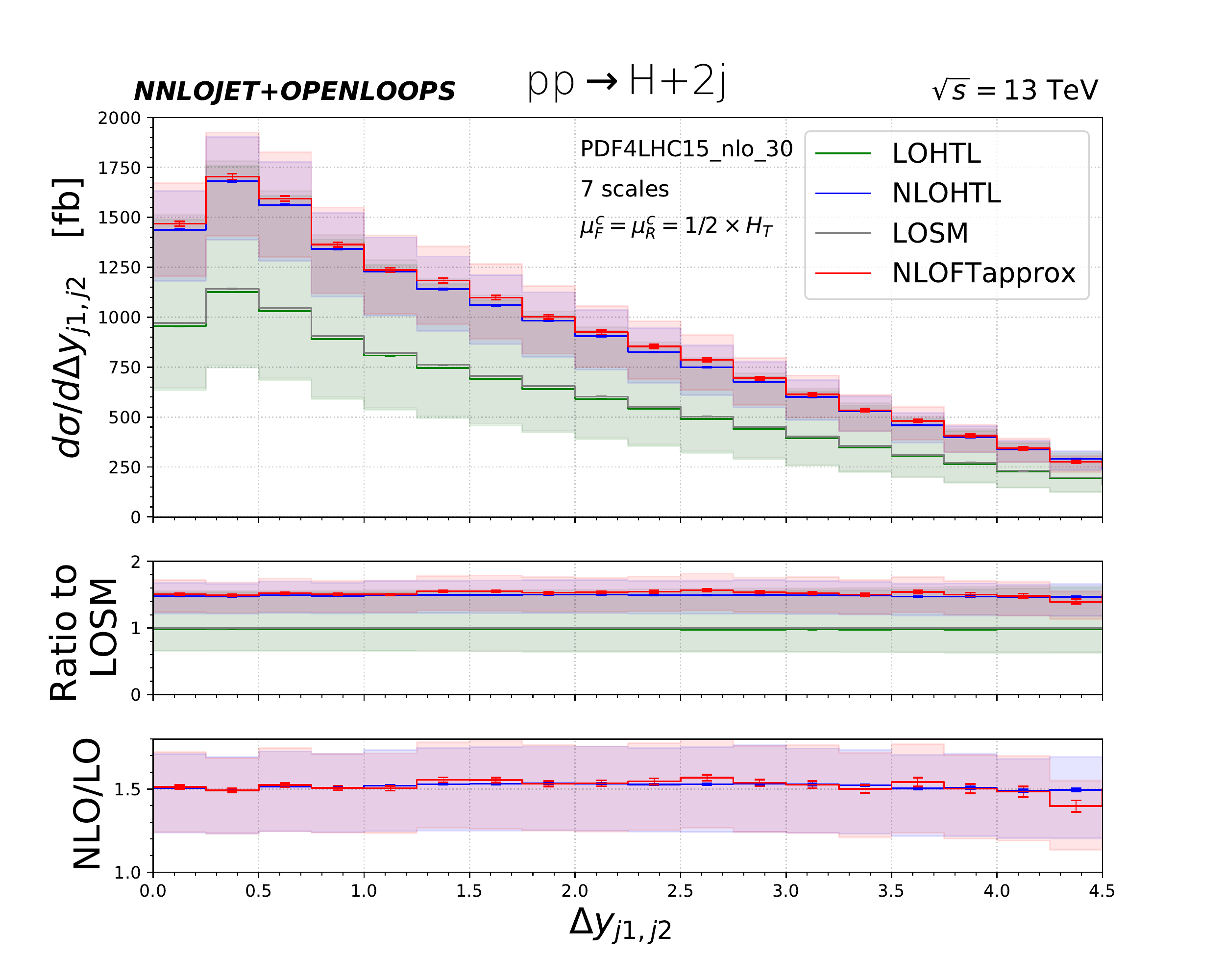}
		\includegraphics[width=0.5\textwidth]{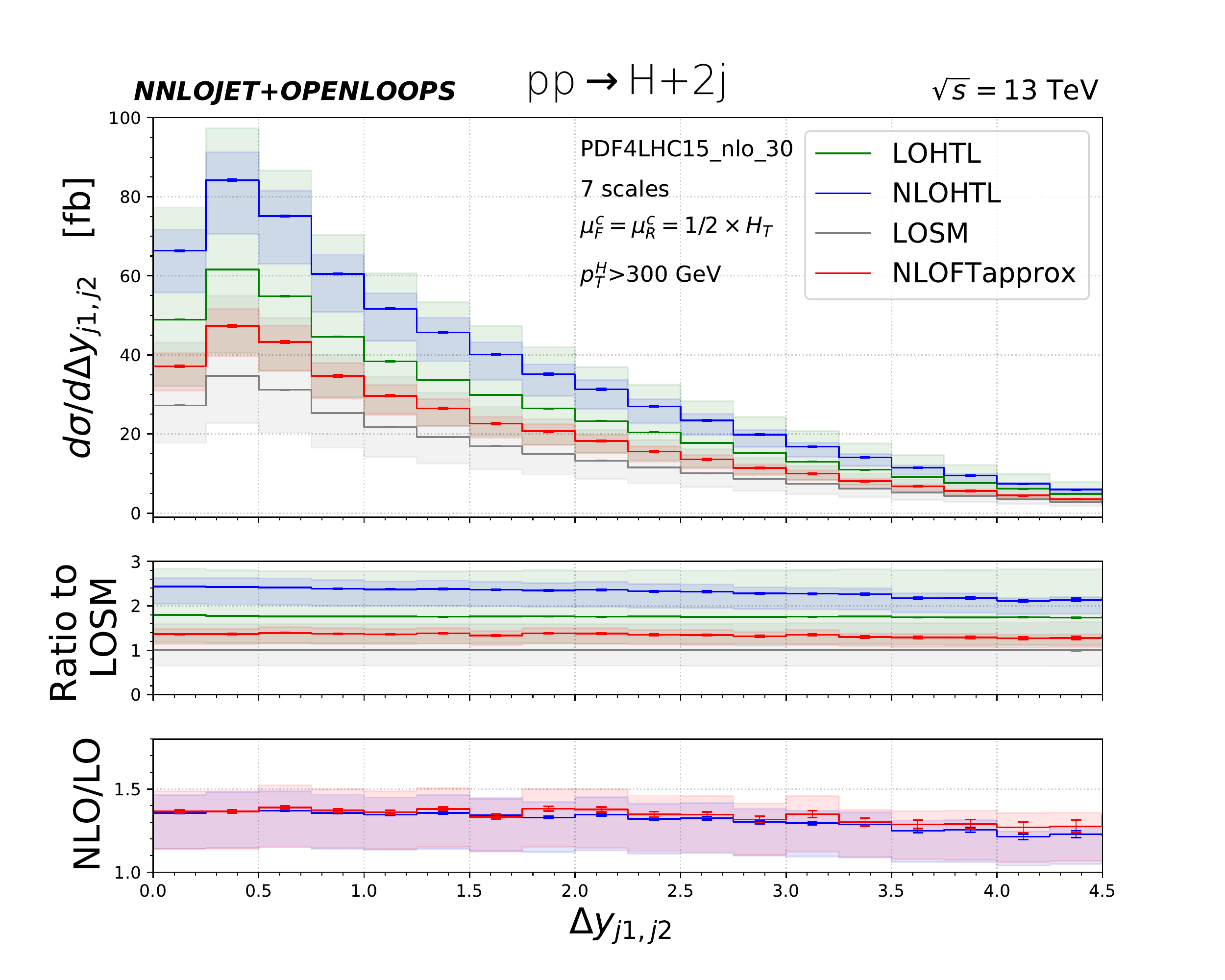}
\caption{Rapidity difference between the two hardest jets in \hjj production. Colour coding and labelling as in Fig.~\ref{fig:hj_pth_ptj}.}
\label{fig:hjj_dyjj}
\end{figure}

In Figs.~\ref{fig:hjj_pth_ptj}-\ref{fig:hjj_dyjj} we turn to the numerical results for \hjj production. We compare NLO corrections in the \HTL with the \FTapprox approximation, focussing on multi-jet observables and jet correlations with and without boosted Higgs kinematics (i.e. inclusive and with an additional \pTHhigh requirement). As discussed in the introduction these are phenomenologically highly relevant for the modelling of \hjj backgrounds in analyses for VBF Higgs production.

In Fig.~\ref{fig:hjj_pth_ptj} we consider the transverse momentum of the Higgs (left) and of the hardest jet (right). These plots can directly be compared with the corresponding ones for \hj production in Fig.~\ref{fig:hj_pth_ptj}. Again we observe very large deviations of the nominal predictions at large transverse momenta. At the same time also the QCD corrections are sizeable: around $60-70\%$ at small \pT 
and around $30\%$ for $\pTH/\pTj = 1$~TeV. However the relative NLO corrections normalised to the respective LO show a universal behaviour, i.e. they are identical in the HTL and the \FTapprox.  

A similar picture as for the transverse momentum distributions emerges when looking at the distribution in  the invariant mass of the Higgs and the hardest jet, as depicted in Fig.~\ref{fig:hjj_mj}. Both, for the inclusive selection (as shown on the left), and for the boosted Higgs selection with an additional \pTHhigh requirement, the QCD corrections in the \FTapprox identically track the corresponding corrections in the HTL, while the nominal predictions substantially diverge.

Next, in Fig.~\ref{fig:hjj_mjj} we turn to the phenomenologically important dijet invariant mass distribution. Again we consider an inclusive selection (left) and a boosted selection requiring \pTHhigh. 
In the inclusive phase-space the NLO corrections are about $50\%$ in the \FTapprox with hardly any  variations over the considered \mjj range. Corrections in the HTL are identical to the \FTapprox at small \mjj and slightly reduce to about $40\%$ in the multi TeV range, i.e. up to $10\%$ smaller than in the \FTapprox. With the boosted selection, corrections in both the \FTapprox and the HTL are at the level of $30-40\%$ and marginally reduce in the tail of the \mjj distribution.

Finally in Fig.~\ref{fig:hjj_dyjj} we plot the rapidity difference between the two jets in \hjj production, again with an inclusive selection on the left and a boosted selection on the right.
In the inclusive case there is hardly any variation in the NLO corrections over the considered rapidity range with a K-factor at the 1.5 level. For the boosted selection, the K-factor decreases slightly from about 1.35 to 1.25 from small to high rapidity differences. For both selections and over the entire rapidity range corrections in the \FTapprox and the HTL agree at the percent level. We observed very similar findings also in other angular correlation observables including e.g. the rapidity difference between the Higgs and the hardest jet.

Overall in all considered observables we find a remarkable agreement of the relative corrections computed in the HTL and the \FTapprox -- despite up to several order of magnitude variations in nominal predictions. This clearly points towards a factorisation of QCD higher-order corrections from the heavy fermion loop mediating the coupling of the Higgs boson.

\section{Conclusions}
\label{sec:conclusions}

In this paper we have presented precise differential predictions for \hj and \hjj production at the LHC at NLO including top-quark mass effects. 
For the former process our prediction incorporates the exact top-quark mass dependence. 
Instead, in our study of \hjj production, the two-loop virtual matrix elements are computed in the HTL approximation (infinite top-quark mass) and reweighed by the full LO result, while the exact top-quark mass dependence is retained in the Born and real radiation contributions.
Our results are produced using the \nnlojet event generator with one-loop
amplitudes provided by \olotter (to be released soon) which implements a novel tensor reduction
method based on the on-the-fly reduction algorithm of OpenLoops.
The two-loop virtual matrix elements including top-quark mass effects contributing to \hj production are evaluated using \secdec.

We find that the inclusion of the exact top-quark mass dependence in the two-loop virtual matrix elements enhances the cross section for \hj production at NLO by about $0.6\%$ with respect to the $\mathrm{FT}_\mathrm{approx}$ prediction, and by about $4.3\%$ with respect to the HTL prediction.
However, the NLO/LO K-factor is found to be universal amongst the three predictions at about $1.7$.
Noteworthy, this universality is broadly also found examining the corrections to the $p_{T,H}$ and $p_{T,j}$ distributions. Although above the top-quark threshold the HTL approximation becomes formally invalid, relative NLO corrections agree at the 10\% (5\%) level between HTL (\FTapprox) respectively and the full theory. 

Regarding the \hjj production process we have produced distributions for the $p_T$ of the Higgs boson and leading jet as well as invariant mass distributions for the Higgs + leading jet system and the leading dijet system. In particular the latter is crucial for gluon-fusion Higgs production backgrounds in VBF Higgs production analyses.
At the inclusive level the $\mathrm{FT}_\mathrm{approx}$ cross section is $2.8\%$ larger than the HTL prediction.
Differentially, we observe that the approximate inclusion of the top-quark mass has the largest, though still rather mild, impact on the $p_T$ distributions above the top-quark threshold. Relative higher-order corrections in the HTL and the $\mathrm{FT}_\mathrm{approx}$ always agree at below the $10\%$ level.


Overall, the top-quark mass effects at NLO are observed to be rather mild. In particular they are expected to be at the same level or even smaller than the current uncertainties due to the scheme dependence of the top-quark mass.

The computations presented are relevant for analyses of Higgs production at large transverse momentum and also for Higgs plus multi-jet backgrounds in Higgs production via vector boson fusion. 
Through detailed comparisons, this study proves the reliability of higher-order corrections computed in the HTL rescaled with lower-order predictions with explicit mass dependence.
The multiplicative reweighting procedures are of similar theoretical uncertainties, within $10\%$, for both event-by-event and bin-by-bin rescaling.
This paves the way towards computations of \pphj including approximate mass effects at NNLO and higher QCD accuracy. 


\section*{Acknowledgments}
We are grateful to Alexander Karlberg for contributions at the early stage of this work. 
We thank Thomas Gehrmann, Gudrun Heinrich, Joey Huston and Stefano Pozzorini for useful discussions and comments on the manuscript.
We thank the University of Zurich S3IT (http://www.s3it.uzh.ch) and Swiss National Supercomputing Centre (CSCS) for providing support and computational resources.
This research is supported in part by the Swiss National Science Foundation (SNF) under contract 200020-175595 and by CSCS under project ID UZH10.
This research was supported in parts by the Deutsche Forschungsgemeinschaft (DFG, German Research Foundation) under grant  396021762 - TRR 257.
J.L. is supported by the Science and Technology Research Council (STFC) under the Consolidated Grant ST/T00102X/1 and the STFC Ernest Rutherford Fellowship ST/S005048/1.
S.P.J. is supported by a Royal Society University Research Fellowship (Grant URF/R1/201268).
J.-N.L. and H.Z. are supported by the Swiss National Science Foundation (SNF)
under contract BSCGI0-157722.

\appendix

\section{Inclusive cross sections }
\label{app:xsections}
In Tab.~\ref{tab:accXSHJ} and Tab.~\ref{tab:accXSHJJ} we list as reference integrated cross sections $\sigma(p_{T,H} > p_{T,H}^{\text{cut}})$ in function of $p_{T,H}^{\text{cut}}$ for \pphj and \pphjj respectively. Shown are cross sections at LO and NLO and related K-factors considering the HTL and \FTapprox approximations, and for \pphj also the results with exact top-mass dependence. The setup is given in Section~\ref{sec:setup}.

\begin{table}[t]
\centering
  \begin{tabular}{  c|ccc |ccccc }
  $p_{T,H}^{\text{cut}}$ & LO$_{\text{HTL}}$ & NLO$_{\text{HTL}}$ & K$^{\text{NLO}}_{\text{HTL}}$ & LO$_{\text{SM}}$ & NLO$_{\FTapprox}$ & K$^{\text{NLO}}_{\FTapprox}$ & NLO$_{\text{SM}}$ & K$^{\text{NLO}}_{\text{SM}}$\\
  \hline
  \hline
50 & $4453^{+1755}_{-1181}$ & $8482^{+1755}_{-1522}$ & $1.90$ & $4566^{+1800}_{-1212}$ & $8682^{+1793}_{-1557}$ & $1.90$ & $8732^{+1858}_{-1585}$ & $1.91$ \\ 
100 & $1430^{+585}_{-389}$ & $2732^{+578}_{-502}$ & $1.91$ & $1391^{+570}_{-379}$ & $2645^{+557}_{-485}$ & $1.90$ & $2669^{+575}_{-494}$ & $1.92$ \\ 
150 & $593^{+249}_{-164}$ & $1121^{+235}_{-207}$ & $1.89$ & $528^{+222}_{-146}$ & $989^{+205}_{-182}$ & $1.87$ & $996^{+216}_{-186}$ & $1.89$ \\ 
200 & $284^{+121}_{-79.6}$ & $533^{+111}_{-98.9}$ & $1.88$ & $219^{+94.2}_{-61.7}$ & $411^{+85.9}_{-76.5}$ & $1.88$ & $417^{+90.0}_{-78.4}$ & $1.90$ \\ 
250 & $151^{+65.2}_{-42.6}$ & $281^{+58.2}_{-52.2}$ & $1.87$ & $97.4^{+42.5}_{-27.7}$ & $184^{+39.3}_{-34.8}$ & $1.89$ & $189^{+41.2}_{-36.1}$ & $1.94$ \\ 
300 & $85.9^{+37.6}_{-24.5}$ & $160^{+32.9}_{-29.7}$ & $1.86$ & $45.9^{+20.3}_{-13.2}$ & $87.8^{+19.2}_{-16.9}$ & $1.91$ & $90.1^{+19.7}_{-17.4}$ & $1.96$ \\ 
350 & $51.8^{+22.9}_{-14.8}$ & $95.9^{+19.7}_{-17.9}$ & $1.85$ & $22.9^{+10.2}_{-6.62}$ & $44.0^{+9.71}_{-8.53}$ & $1.92$ & $45.1^{+10.2}_{-8.85}$ & $1.97$ \\ 
400 & $32.5^{+14.5}_{-9.38}$ & $60.1^{+12.3}_{-11.2}$ & $1.85$ & $12.0^{+5.39}_{-3.48}$ & $23.0^{+5.10}_{-4.49}$ & $1.92$ & $23.6^{+5.38}_{-4.66}$ & $1.98$ \\ 
450 & $21.2^{+9.49}_{-6.14}$ & $39.1^{+8.00}_{-7.33}$ & $1.84$ & $6.52^{+2.96}_{-1.91}$ & $12.6^{+2.80}_{-2.46}$ & $1.93$ & $12.9^{+2.97}_{-2.56}$ & $1.98$ \\ 
500 & $14.2^{+6.39}_{-4.13}$ & $26.2^{+5.36}_{-4.92}$ & $1.84$ & $3.67^{+1.68}_{-1.08}$ & $7.09^{+1.59}_{-1.40}$ & $1.93$ & $7.25^{+1.69}_{-1.45}$ & $1.97$ \\ 
550 & $9.71^{+4.40}_{-2.84}$ & $17.9^{+3.66}_{-3.37}$ & $1.84$ & $2.14^{+0.98}_{-0.63}$ & $4.11^{+0.92}_{-0.81}$ & $1.92$ & $4.18^{+0.96}_{-0.83}$ & $1.96$ \\ 
600 & $6.79^{+3.09}_{-1.99}$ & $12.5^{+2.56}_{-2.36}$ & $1.84$ & $1.28^{+0.59}_{-0.38}$ & $2.47^{+0.56}_{-0.49}$ & $1.93$ & $2.50^{+0.59}_{-0.51}$ & $1.96$ \\ 
650 & $4.82^{+2.21}_{-1.42}$ & $8.88^{+1.83}_{-1.69}$ & $1.84$ & $0.78^{+0.36}_{-0.23}$ & $1.50^{+0.34}_{-0.30}$ & $1.92$ & $1.53^{+0.36}_{-0.31}$ & $1.96$ \\ 
700 & $3.48^{+1.60}_{-1.03}$ & $6.42^{+1.33}_{-1.22}$ & $1.85$ & $0.49^{+0.23}_{-0.15}$ & $0.94^{+0.21}_{-0.19}$ & $1.92$ & $0.96^{+0.23}_{-0.19}$ & $1.96$ \\ 
750 & $2.54^{+1.17}_{-0.75}$ & $4.68^{+0.97}_{-0.89}$ & $1.84$ & $0.31^{+0.15}_{-0.09}$ & $0.60^{+0.14}_{-0.12}$ & $1.94$ & $0.62^{+0.14}_{-0.13}$ & $1.97$ \\ 
800 & $1.87^{+0.87}_{-0.56}$ & $3.45^{+0.71}_{-0.66}$ & $1.84$ & $0.20^{+0.1}_{-0.06}$ & $0.39^{+0.09}_{-0.08}$ & $1.92$ & $0.39^{+0.09}_{-0.08}$ & $1.94$
  \end{tabular}
  \caption{\label{tab:accXSHJ}Integrated cross sections in fb depending on $p_{T,H}^{\text{cut}}$ at LO and NLO in the HTL, \FTapprox, and with full top-quark mass dependence (SM) for \hj production at the LHC with $\sqrt{S}=13$\,TeV together with corresponding K-factors. Uncertainties correspond to the envelope of 7-point scale variations. We require $p_{\rm{T},j}>30\,\GeV$. }
\end{table}

\begin{table}[t]
\centering
  \begin{tabular}{  l|ccc |ccc }
  $p_{T,H}^{\text{cut}}$ & LO$_{\text{HTL}}$ & NLO$_{\text{HTL}}$ & K$^{\text{NLO}}_{\text{HTL}}$ & LO$_{\text{SM}}$ & NLO$_{\FTapprox}$ & K$^{\text{NLO}}_{\FTapprox}$\\
  \hline
  \hline
50 & $2365^{+1377}_{-813}$ & $3511^{+446}_{-631}$ & $1.48$ & $2387^{+1391}_{-821}$ & $3574^{+477}_{-651}$ & $1.50$ \\ 
100 & $1355^{+790}_{-466}$ & $2025^{+267}_{-368}$ & $1.50$ & $1317^{+769}_{-454}$ & $1991^{+276}_{-368}$ & $1.51$ \\ 
150 & $671^{+392}_{-232}$ & $969^{+110}_{-169}$ & $1.44$ & $601^{+353}_{-208}$ & $881^{+107}_{-157}$ & $1.47$ \\ 
200 & $355^{+208}_{-123}$ & $497^{+47.9}_{-83.4}$ & $1.40$ & $281^{+166}_{-97.5}$ & $399^{+42.1}_{-68.7}$ & $1.42$ \\ 
250 & $201^{+118}_{-69.6}$ & $274^{+22.7}_{-44.4}$ & $1.36$ & $136^{+80.2}_{-47.2}$ & $188^{+17.8}_{-31.6}$ & $1.39$ \\ 
300 & $120^{+70.5}_{-41.6}$ & $160^{+11.8}_{-25.3}$ & $1.34$ & $68.1^{+40.3}_{-23.7}$ & $92.4^{+7.73}_{-15.2}$ & $1.36$ \\ 
350 & $74.8^{+44.0}_{-26.0}$ & $98.5^{+6.53}_{-15.2}$ & $1.32$ & $35.5^{+21.1}_{-12.4}$ & $47.7^{+3.70}_{-7.73}$ & $1.34$ \\ 
400 & $48.4^{+28.5}_{-16.8}$ & $62.9^{+3.82}_{-9.56}$ & $1.30$ & $19.2^{+11.4}_{-6.71}$ & $25.4^{+1.82}_{-4.05}$ & $1.33$ \\ 
450 & $32.2^{+19.0}_{-11.2}$ & $41.5^{+2.36}_{-6.23}$ & $1.29$ & $10.7^{+6.38}_{-3.75}$ & $14.1^{+0.96}_{-2.23}$ & $1.32$ \\ 
500 & $21.9^{+12.9}_{-7.62}$ & $28.0^{+1.51}_{-4.18}$ & $1.28$ & $6.16^{+3.68}_{-2.16}$ & $8.06^{+0.52}_{-1.27}$ & $1.31$ \\ 
550 & $15.2^{+8.99}_{-5.30}$ & $19.4^{+1.01}_{-2.87}$ & $1.27$ & $3.64^{+2.18}_{-1.28}$ & $4.74^{+0.30}_{-0.74}$ & $1.30$ \\ 
600 & $10.8^{+6.37}_{-3.75}$ & $13.7^{+0.69}_{-2.02}$ & $1.27$ & $2.20^{+1.32}_{-0.77}$ & $2.85^{+0.18}_{-0.44}$ & $1.29$ \\ 
650 & $7.72^{+4.57}_{-2.70}$ & $9.79^{+0.49}_{-1.44}$ & $1.27$ & $1.36^{+0.82}_{-0.48}$ & $1.76^{+0.11}_{-0.27}$ & $1.29$ \\ 
700 & $5.61^{+3.33}_{-1.96}$ & $7.10^{+0.35}_{-1.05}$ & $1.27$ & $0.85^{+0.51}_{-0.30}$ & $1.10^{+0.07}_{-0.17}$ & $1.29$ \\ 
750 & $4.12^{+2.45}_{-1.44}$ & $5.22^{+0.25}_{-0.77}$ & $1.26$ & $0.55^{+0.33}_{-0.19}$ & $0.70^{+0.04}_{-0.11}$ & $1.29$ \\ 
800 & $3.06^{+1.82}_{-1.07}$ & $3.88^{+0.19}_{-0.57}$ & $1.27$ & $0.36^{+0.22}_{-0.13}$ & $0.46^{+0.03}_{-0.07}$ & $1.28$
  \end{tabular}
  \caption{\label{tab:accXSHJJ}Integrated cross sections in fb depending on $p_{T,H}^{\text{cut}}$ at LO and NLO in the HTL and \FTapprox for \hjj production at the LHC with $\sqrt{S}=13$\,TeV together with corresponding K-factors. Uncertainties correspond to the envelope of 7-point scale variations. We require $p_{\rm{T},j_1}>40\,\GeV,\,$ $p_{\rm{T},j_2}>30\,\GeV$.}
\end{table}

\clearpage

\bibliographystyle{JHEP}

\bibliography{hjsm.bib}

\end{document}